\documentclass[prl,twocolumn,superscriptaddress,nobibnotes,a4paper]{revtex4-2}
\usepackage[pdftex]{graphicx}
\usepackage[pdftex]{epsfig}
\usepackage{amsmath}
\usepackage{amssymb}
\usepackage{amsfonts}
\usepackage{color}
\usepackage{wrapfig}
\usepackage{eucal}
\usepackage{hhline}
\usepackage{threeparttable}
\usepackage{siunitx}
\usepackage{supertabular}
\usepackage[normalem]{ulem}
\usepackage{xcolor}
\usepackage{multirow}
\usepackage{tabularx}
\usepackage{upgreek}
\usepackage{float}
\usepackage{soul}

\usepackage[export]{adjustbox}
\usepackage{soul}
\usepackage[version=4]{mhchem} 
\usepackage{array} 
\usepackage[colorlinks=true, allcolors=blue]{hyperref} 
\usepackage{comment}

\setstcolor{red}
\graphicspath{ {./fig_pdf/} }
\usepackage{graphicx}

\begin{document}
\setstcolor{red}
\title{Efficient Conversion of Optical to Mechanical States\\Close to the Single-Quantum Level}

\author{Alexander Rolf Korsch}
\thanks{These authors contributed equally to this work}
\affiliation{Kavli Institute of Nanoscience, Department of Quantum Nanoscience, Delft University of Technology, 2628CJ Delft, The Netherlands}

\author{Liu Chen}
\thanks{These authors contributed equally to this work}
\affiliation{Kavli Institute of Nanoscience, Department of Quantum Nanoscience, Delft University of Technology, 2628CJ Delft, The Netherlands}

\author{Pedro V. Pinho}
\thanks{These authors contributed equally to this work}
\affiliation{Kavli Institute of Nanoscience, Department of Quantum Nanoscience, Delft University of Technology, 2628CJ Delft, The Netherlands}
\affiliation{Instituto de F\'isica Gleb Wataghin, Universidade Estadual de Campinas (UNICAMP), 13083-859 Campinas, SP, Brazil}

\author{Boris Müllendorff}
\affiliation{Kavli Institute of Nanoscience, Department of Quantum Nanoscience, Delft University of Technology, 2628CJ Delft, The Netherlands}

\author{Jan N. Kirchhof}
\affiliation{Kavli Institute of Nanoscience, Department of Quantum Nanoscience, Delft University of Technology, 2628CJ Delft, The Netherlands}

\author{Yong Yu}
\affiliation{Kavli Institute of Nanoscience, Department of Quantum Nanoscience, Delft University of Technology, 2628CJ Delft, The Netherlands}

\author{Thiago P. Mayer Alegre}
\affiliation{Instituto de F\'isica Gleb Wataghin, Universidade Estadual de Campinas (UNICAMP), 13083-859 Campinas, SP, Brazil}

\author{Simon Gr\"oblacher}
\email{s.groeblacher@tudelft.nl}
\affiliation{Kavli Institute of Nanoscience, Department of Quantum Nanoscience, Delft University of Technology, 2628CJ Delft, The Netherlands}

\begin{abstract}
    Coherent interfaces between optical photons and mechanical excitations provide a promising route towards phonon-state engineering and hybrid quantum information processing. Cavity optomechanical systems enable such interfaces via optomechanically induced transparency (OMIT), allowing coherent mapping between traveling optical fields and localized mechanical modes. However, previous implementations of OMIT conversion protocols were limited to classical input signals with large coherent state photon numbers due to room temperature operation and corresponding thermal mechanical noise. Here, we demonstrate efficient low-noise photon–phonon state transfer close to the single-quantum regime in an optomechanical crystal operated at Millikelvin temperatures. Using weak coherent optical input pulses at the few-photon level, we achieve a record-level photon–phonon conversion efficiency of $\eta=0.76$, a mechanical storage lifetime of $T_\mathrm{1}=\SI{7.3}{\micro\second}$, and a tunable conversion bandwidth exceeding $\SI{4.5}{\mega\hertz}$. Hanbury Brown–Twiss measurements of the retrieved signal demonstrate the coherent nature of the converted phononic state, evidencing low added thermal noise in the conversion process ($n_\mathrm{th}=9.0$). These results establish optomechanical crystals as efficient optical interfaces to GHz mechanical modes and provide a pathway toward deterministic single-quantum-level mechanical state preparation.
\end{abstract}
\newpage
\maketitle
\section*{Introduction}
    Non-classical states of mechanical motion are a versatile resource for quantum science and technologies~\cite{barzanjeh_optomechanics_2022}. Quantum states of massive mechanical systems enable fundamental tests of quantum mechanics by probing decoherence mechanisms and constraining collapse models~\cite{bose_massive_2025}. Moreover, their long coherence times and ability to couple to a wide range of other solid-state qubit systems position mechanical quantum states as a promising platform for quantum metrology and quantum information processing~\cite{aspelmeyer_cavity_2014}. Despite these prospects, the deterministic generation of non-classical phonon states remains a significant experimental challenge. To date, most demonstrations rely on heralded schemes, such as the preparation of single-phonon Fock states via optomechanical interactions combined with single-photon detection~\cite{riedinger_non-classical_2016, hong_hanbury_2017}. Alternatively, non-classical mechanical states have been engineered through coupling to intrinsically non-linear quantum systems, such as superconducting qubits~\cite{bild_schrodinger_2023, marti_quantum_2024, satzinger_quantum_2018, ma_non-classical_2021}.

    In contrast, the generation and manipulation of non-classical states of light are well established. Nonlinear optical systems and quantum emitters routinely enable the production of squeezed states~\cite{walls_squeezed_1983, schnabel_squeezed_2017}, single-~\cite{tomm_bright_2021} and multi-photon Fock states~\cite{sonoyama_generation_2024, cooper_experimental_2013}, as well as states exhibiting complex phase-space structures, such as Schrödinger cat states~\cite{simon_experimental_2024, sychev_enlargement_2017, takahashi_generation_2008, ourjoumtsev_generation_2007, neergaard-nielsen_generation_2006} or Gottesman–Kitaev–Preskill states~\cite{larsen_integrated_2025}. Coherent interfaces that allow to map itinerant optical quantum states onto mechanical excitations, therefore, provide a powerful strategy for phonon-state engineering by photon-to-phonon conversion. Cavity optomechanical systems offer a particularly promising platform for implementing such interfaces~\cite{aspelmeyer_cavity_2014}: via the radiation-pressure interaction, optical fields can be coherently coupled to long-lived mechanical modes whose resonance frequency, linewidth, and spatial confinement can be engineered over wide parameter ranges. This versatility facilitates not only the storage of optical quantum states in mechanical modes, enabling quantum memory applications, but more generally coherent photon–phonon state transfer for preparing complex phononic quantum states.

\begin{figure}
    	\includegraphics[width = 1.0 \linewidth]{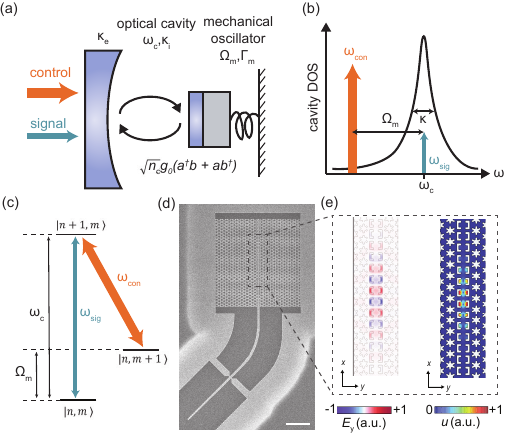} 
    	\caption{\textbf{Photon--phonon conversion via OMIT in optomechanical crystals.} (a) Schematic illustration of a cavity optomechanical system with optical (mechanical) resonance frequency $\omega_\mathrm{c}$ ($\Omega_\mathrm{m}$) and loss rates $\kappa$ ($\Gamma_\mathrm{m}$). The optical loss rate $\kappa$ is composed of intrinsic cavity losses $\kappa_\mathrm{i}$ and extrinsic cavity coupling $\kappa_\mathrm{e}$ to an input mode. A strong control beam detuned to the red-optomechanical sideband with laser detuning $\Delta=\omega_\mathrm{con}-\omega_\mathrm{c}=-\Omega_\mathrm{m}$ and laser frequency $\omega_\mathrm{con}$ drives the optical mode $\hat{a}$ with an intracavity photon number $n_\mathrm{c}$. This driving enables coherent photon--phonon conversion of an optical input signal field at frequency $\omega_\mathrm{sig}\approx\omega_\mathrm{c}$ onto the mechanical oscillator mode $\hat{b}$ through optomechanically-induced transparency (OMIT) with single-photon optomechanical coupling rate $g_0$. (b) Control and signal driving scheme and (c) energy level scheme of the coupled optomechanical system. (d) Scanning electron microscope image of the quasi-2D optomechanical crystal device. The scale bar corresponds to $\qty{5}{\micro m}$. (e) Finite-element simulation of the $y$-component of the electromagnetic field $E_y$ of the optical mode (left) and of the displacement $u=|\vec{u}|$ of the mechanical mode (right).}
    	\label{Fig1_concept}
    \end{figure}

    One implementation of such a photon--phonon interface in cavity optomechanics is optomechanically induced transparency (OMIT). OMIT arises when a weak signal field interrogates an optical cavity with resonance frequency $\omega_\mathrm{c}$ and linewidth $\kappa$ that is simultaneously driven by a strong control field at frequency $\omega_\mathrm{con}$ red-detuned from the optical resonance by one mechanical frequency $\Omega_\mathrm{m}$, such that $\omega_{\text{con}} - \omega_{\text{c}} = -\Omega_{\text{m}}$ as illustrated in Fig.~\ref{Fig1_concept}(a) and (b)~\cite{weis_optomechanically_2010,safavi-naeini_electromagnetically_2011}. In the resolved-sideband regime ($\Omega_{\text{m}} \gg \kappa$), the optomechanical interaction enables coherent exchange of excitations between the optical ($\hat{a}$) and mechanical ($\hat{b}$) modes through an effective beam-splitter Hamiltonian $H_{\text{int}} = \hbar G (\hat{a}^\dagger \hat{b} + \hat{a} \hat{b}^\dagger)$ with an enhanced coupling rate $G = \sqrt{n_{\text{c}}}\, g_0$ set by the intracavity control-field photon number $n_{\text{c}}$ and the optomechanical single-photon coupling strength $g_0$. The interference responsible for OMIT originates from two pathways:\ (1) the signal field directly excites the optical transition $\lvert n,m\rangle \rightarrow \lvert n+1,m\rangle$ with optical (mechanical) mode occupation $n$ ($m$); (2) simultaneously, the beating between the signal and control fields coherently drives the mechanical mode, enabling a second, mechanically mediated up-conversion process $\lvert n,m+1\rangle \rightarrow \lvert n+1,m\rangle$ (see Fig.~\ref{Fig1_concept}(c)). Destructive interference between these two excitation pathways suppresses signal absorption and opens a narrow transparency window within the optical resonance. Furthermore, the interference between the signal and control field allows the temporal and phase information of the signal to be mapped directly onto the mechanical mode. The maximum achievable photon-phonon conversion efficiency $\eta=\eta_\mathrm{c}$ is determined by the external coupling efficiency of the optical cavity mode $\eta_\mathrm{c}=\kappa_\mathrm{e}/\kappa$. By switching off the control field immediately after the conversion process, the optomechanical coupling is effectively quenched and the mechanical excitation is stored for the intrinsic lifetime of the mechanical oscillator. Re-applying the red-detuned control field, enables coherent phonon-to-photon conversion via optomechanical anti-Stokes scattering and retrieval of the optical signal. OMIT-based interfaces have been realized in silica microspheres~\cite{fiore2011storing}, diamond microdisks~\cite{lake2021processing}, as well as soft-clamped silicon nitride membranes~\cite{kristensen_2024}. However, due to room-temperature operation and the resulting thermal mechanical occupancy ranging from $n_\mathrm{th}=10^3$ to $10^6$, these previous implementations have been limited to classical optical input states with macroscopic photon numbers.
    
    Here, we demonstrate a low-noise photon-phonon interface based on a quasi-two-dimensional optomechanical crystal (OMC) at Millikelvin temperatures that operates close to the single-quantum level. The device enables coherent conversion between traveling optical fields and localized mechanical excitations with high efficiency ($\eta=0.76$) and large bandwidth up to $\SI{4.5}{\mega\hertz}$, allowing reversible mapping of few-photon-level optical pulses while maintaining low added thermal noise of $n_\mathrm{th}=9.0$. To characterize the photon-phonon interface, we probe the device with weak coherent-state pulses on the few-photon level and verify the coherent nature of the converted mechanical state through Hanbury Brown-Twiss measurements of the phonon autocorrelation function. While the generated phononic states in our experiment remain classical, our method allows quantitative evaluation of conversion efficiency, bandwidth, and added thermal noise \textemdash crucial quantities for future operation with single-photon or other non-classical optical states. To date, single-excitation operation in optomechanical systems has only been accessible via post-selection, owing to the relatively small optomechanical coupling rates and the need for strong drive fields that inevitably populate large thermal states. With improved thermal anchoring and reduced optical absorption, this platform provides a direct path toward single-phonon-level operation, enabling controlled preparation and characterization of non-classical phononic states.
\section*{Methods}
    The quasi-two-dimensional suspended OMC cavity was fabricated on a silicon-on-insulator (SOI) platform, as shown in Fig.~\ref{Fig1_concept}(d). The two-dimensional anchoring configuration provides an efficient pathway for the dissipation of thermal phonons generated through optical absorption, thereby minimizing the temperature rise within the structure during optical measurements~\cite{Ren2020}. Both the optical and mechanical modes of the device (see Fig.~\ref{Fig1_concept}(e)) are confined along the y-axis by a snowflake-patterned phoxonic band gap crystal~\cite{Amir-PRL, Kersul-2023}. Confinement along the x-axis is achieved through a periodic array of C-shaped holes with spatially modulated dimensions, forming the optical and mechanical cavity region. The strong spatial overlap of the optical and mechanical modes within the small cavity volume leads to significant optomechanical coupling, with a simulated single-photon coupling rate of $g_{0}/2\pi = \SI{1.0}{\mega\hertz}$ and mechanical frequency $\Omega_\mathrm{m}/2\pi = \SI{10.314}{\giga\hertz}$. To maximize OMIT conversion efficiency, we choose to work with an overcoupled optical cavity mode with linewidth $\kappa/2\pi = \SI{6.43}{\giga\hertz}$ and external coupling efficiency $\kappa_\mathrm{e}/\kappa = 0.76$, realized by adequate design of the optical waveguide-cavity coupler.

\begin{figure}
    	\includegraphics[width = 1.0 \linewidth]{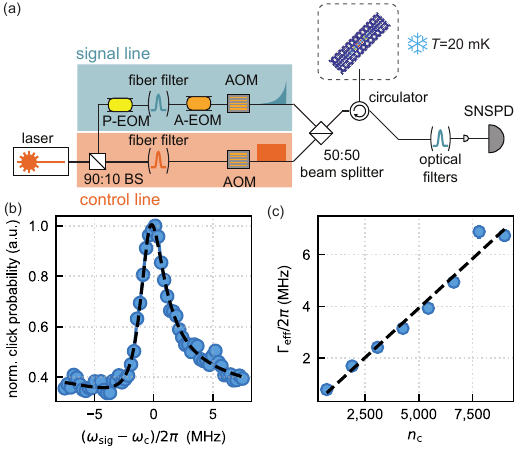} 
    	\caption{\textbf{Experimental setup and OMIT window measurement.} (a) Experimental setup for OMIT measurement. Laser light is split into two optical paths: the control line and the signal line. The signal line pulse is shaped by an amplitude-electro-optic modulator (A-EOM) into an exponentially rising pulse with bandwidth $\Gamma_\mathrm{sig}$. BS, beam splitter; P-EOM, phase electro-optic modulator; AOM, acousto-optic modulator; SNSPD, superconducting nanowire single photon detector. (b) Measured OMIT window when sweeping the detuning between signal light $\omega_\mathrm{sig}$ and cavity resonance $\omega_\mathrm{c}$. (c) Dependence of the OMIT linewidth $\Gamma_\mathrm{eff}$ on intracavity photon number $n_\mathrm{c}$. The linear fit gives a single photon optomechanical coupling strength $g_0/2\pi = \SI{1.10}{MHz}$. This dependence on $n_\mathrm{c}$ demonstrates an achievable bandwidth of more than $\SI{6}{MHz}$.}
    	\label{Fig2_setup and OMIT}
    \end{figure}

    The experimental setup consists of two optical paths, labeled the control and signal lines as indicated in Fig.~\ref{Fig2_setup and OMIT}(a). Both lines are derived from a single laser locked to the red optomechanical sideband of the OMC device at detuning $\Delta = -\Omega_\mathrm{m}$ whose light is split via a 90:10 beam splitter. The laser light on the control line is filtered using a fiber-coupled Fabry-P\'erot cavity to remove laser phase noise and shaped into rectangular pulses of length $T_\mathrm{con}$ and intracavity photon number $n_\mathrm{c}$ by an acousto-optic modulator (AOM). On the signal line, a phase electro-optic modulator (P-EOM) generates sidebands at the mechanical frequency $\pm \Omega_\mathrm{m}$. A fiber filter locked to the blue EOM sideband filters out the carrier and red EOM sideband to transmit only a single signal frequency, which is resonant with the optical cavity of the OMC device ($\omega_\mathrm{sig}=\omega_\mathrm{c}$). To match the bandwidth of the input signal to the OMIT window, the signal pulse is shaped using an amplitude EOM. To increase the pulse extinction ratio, the signal pulse is further truncated using an AOM gated by a rectangular pulse of length $T_\mathrm{sig}$. The control pulse and signal pulse are sent simultaneously to the device inside a dilution refrigerator at $T=\SI{20}{mK}$. Optical filters in the detection path locked at the OMC cavity resonance frequency are used to reject the control pulse, enabling the measurement of the transmitted light on a superconducting nanowire single photon detector (SNSPD).

    We characterize the OMIT window of the OMC device by sweeping the detuning between the signal light frequency $\omega_\mathrm{sig}$ and optical cavity resonance $\omega_\mathrm{c}$. To properly resolve the OMIT window, we employ a long control pulse of length $T_\mathrm{con}=\SI{4}{\mu s}$, ensuring that the spectral bandwidth remains narrower than the OMIT window's bandwidth $\Gamma_\mathrm{eff} = \Gamma_\mathrm{m} + \Gamma_\mathrm{opt}$ where $\Gamma_\mathrm{m}$ is the intrinsic mechanical damping rate and $\Gamma_\mathrm{opt} = 4g_0^2 n_\mathrm{c}/\kappa$ is the optomechanically induced damping rate. The signal pulse is an exponentially rising waveform with $\Gamma_\mathrm{sig}/2\pi=\SI{53}{kHz}$ and $T_\mathrm{sig}=\SI{4}{\micro\second}$. Sweeping the signal frequency across $\omega_\mathrm{c}$ reveals a Fano-shaped transparency window, from which we extract the OMIT bandwidth, as shown in Fig.~\ref{Fig2_setup and OMIT}(b). By varying the control pulse intracavity photon number $n_\mathrm{c}$, we observe a linear increase of the OMIT bandwidth $\Gamma_\mathrm{eff}$ as shown in Fig.~\ref{Fig2_setup and OMIT}(c). Fitting this dependence yields a single-photon optomechanical coupling strength of $g_0/2\pi = \SI{1.10}{\mega\hertz}$. In addition, we independently calibrate $g_0$ from measurements of the optomechanically scattered photon rate finding $g_0/2\pi=\SI{0.97}{\mega\hertz}$ in good agreement with the value obtained from the OMIT measurements and simulations (see Supplementary Information).

\section*{Results}
    After the initial characterization of our device, we proceed to implement the OMIT photon-phonon conversion protocol described above using weak coherent signal pulses at the few-photon level. We first characterize the dependence of the OMIT conversion efficiency on the bandwidth matching between the signal pulse and the OMIT window. Subsequently, we assess the role of thermal noise in the OMIT conversion process by benchmarking the thermal occupancy of the mechanical mode and investigate its dependence on the length of the control pulse. Lastly, we measure the lifetime of the converted phonons and total efficiency upon retrieval.

\begin{figure}
    	\includegraphics[width = 1.0 \linewidth]{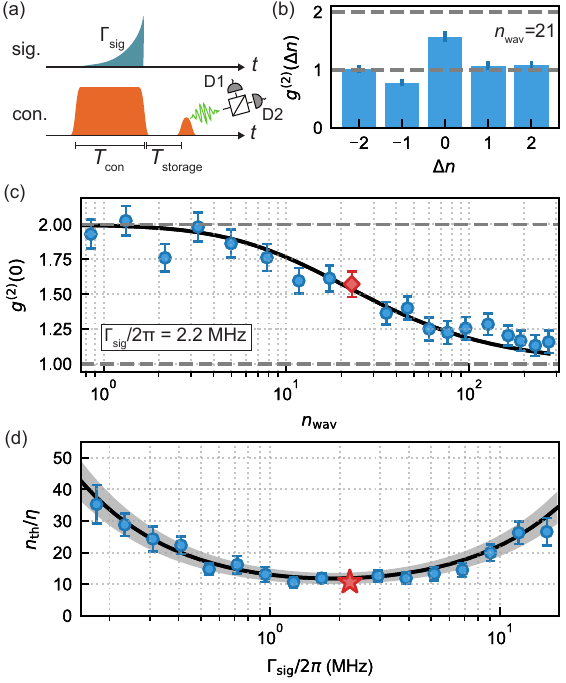} 
    	\caption{\textbf{Hanbury Brown-Twiss measurement of the mechanical state generated via photon--phonon conversion.} (a) Pulse sequence for OMIT conversion and readout. A strong control pulse of length $T_\mathrm{con}=\qty{2.4}{\micro s}$ and intracavity photon number $n_\mathrm{c}=3,200$ ($\Gamma_\mathrm{eff}/2\pi=\SI{1.9}{\mega\hertz}$) enables coherent conversion of an exponentially rising weak coherent input signal with bandwidth $\Gamma_\mathrm{sig}$ and $T_\mathrm{sig}=\SI{2.4}{\micro\second}$. After a delay $T_\mathrm{storage}=\SI{150}{\nano\second}$, a second short control pulse reconverts part of the mechanical excitation back into optical photons (see Supplementary Information for experimental details). The retrieved optical signal is sent to a Hanbury Brown-Twiss (HBT) setup. (b) Second order autocorrelation function $g^{(2)}(\Delta n)$ of detection events on detectors D1 and D2 in the HBT measurement shifted by $\Delta n$ repetitions of the pulse sequence. (c) Measured $g^{(2)}(0)$ as a function of input signal photon number coupled into the device waveguide $n_\mathrm{wav}$. The solid line is a fit to Eq.~\eqref{eq_g2_fit} to extract the input-referred thermal noise of the OMIT conversion process $n_\mathrm{th}/\eta$. The red diamond indicates the data point corresponding to the measurement shown in (b). Dashed grey lines in (b) and (c) show the expected value of the autocorrelation function for a thermal ($g^{(2)}_\mathrm{th}(0)=2$) and coherent state ($g^{(2)}_\mathrm{coh}(0)=1$). (d) Input-referred thermal noise measured as a function of signal bandwidth $\Gamma_\mathrm{sig}$. The solid line corresponds to the theoretical model in Eq.~\eqref{eq_bandwidth_matching_theory}, where the added thermal noise $n_\mathrm{th}=9.0$ from the strong control pulse is independently calibrated (see Fig.~\ref{Fig4_control_pulse_length}(a)) and the shaded area shows the uncertainty of the theoretical model due to a relative error of $15\%$ in the calibration of $n_\mathrm{th}$. The red star indicates the data point corresponding to the measurement shown in (c). All error bars represent one standard deviation calculated from the binomial statistics of count events during the HBT measurement.}
        \label{Fig3_pulse_g2_and_bandwidth}
    \end{figure}

    We use the OMIT window created by the strong red-detuned control pulse to coherently map the weak coherent input signal onto mechanical excitations via photon-phonon conversion by employing the pulse sequence shown in Fig.~\ref{Fig3_pulse_g2_and_bandwidth}(a). Efficient conversion requires the spectral shape of the input pulse to match the OMIT window bandwidth $\Gamma_\mathrm{eff}$~\cite{kristensen_2024}, which we achieve by shaping it into an exponentially rising wave packet with bandwidth $\Gamma_\mathrm{sig}$, mirroring protocols based on electromagnetically induced transparency where optimal storage arises from temporal-mode matching to the transparency window~\cite{Gorshkov2007, Novikova2008}. This condition is conceptually related to virtual coupling, where complete energy transfer can be achieved via tailoring of the transient excitation of the cavity mode~\cite{Hinney2024VirtualCriticalCoupling, Radi2020VirtualCriticalCoupling}. After a controlled delay following the conversion step $T_\mathrm{storage}$, we use a second, weak red-detuned optical pulse to swap a small part of the stored mechanical excitation back onto optical photons through the optomechanical beam-splitter interaction. Hanbury Brown-Twiss (HBT) measurements on the retrieved optical photons allow us to determine the second-order autocorrelation $g^{(2)}(\Delta n)$ of the converted mechanical excitation~\cite{hong_hanbury_2017}, where $\Delta n$ denotes the separation between detection events on detectors D1 and D2 in units of pulse sequence repetitions (see Fig.~\ref{Fig3_pulse_g2_and_bandwidth}(b)). The optomechanical scattering probability of the readout pulse is chosen to be small $p_\mathrm{aS}^\mathrm{(r)}\ll 1$ to minimize additional added thermal noise from the readout process.
    
    Measurements of the phonon statistics of the converted mechanical excitation through $g^{(2)}(0)$ enable us to distinguish the coherent fraction of the converted phonons from thermal noise primarily induced by optical absorption heating from the strong control pulse. Conversion of weak coherent signal photons prepares the mechanical mode in a weak coherent state with an expected value of the autocorrelation function of $g^{(2)}(0)=1$. However, optical absorption heating introduces a thermal component in the mechanical state with occupancy $n_\mathrm{th}$ and expected $g^{(2)}(0)=2$. To separate the two contributions, we measure $g^{(2)}(0)$ as a function of the signal photon number coupled into the device waveguide $n_\mathrm{wav}$ which constitutes the input signal of the OMIT conversion process. To ensure sufficiently low click probabilities for reliable determination of the autocorrelation function, we vary $p_\mathrm{aS}^\mathrm{(r)}$ depending on the number of signal photons $n_\mathrm{wav}$ (see Supplementary Information). As shown in Fig.~\ref{Fig3_pulse_g2_and_bandwidth}(c), increasing $n_\mathrm{wav}$ while keeping $\Gamma_\mathrm{sig}/2\pi$ ($\Gamma_\mathrm{eff}/2\pi$) fixed at $\SI{2.2}{MHz}$ ($\SI{1.9}{MHz}$) leads to a larger fraction of coherently stored phonons in the mechanical mode and thus causes a reduction of $g^{(2)}(0)$. Even for input signals with less than 10 photons, we observe a significant reduction in $g^{(2)}(0)$, demonstrating the capability of our system to perform low-noise photon-phonon conversion at close-to single-quantum signal levels. To extract the input-referred thermal noise $n_\mathrm{th}/\eta$ of our interface, where $\eta$ is the conversion efficiency, we fit the results in Fig.~\ref{Fig3_pulse_g2_and_bandwidth}(c) to a displaced thermal state model derived in the Supplementary Information~\cite{forsch_microwave--optics_2020}
    \begin{align}
        \label{eq_g2_fit}
        g^{(2)}(0)&=2 -\left(\frac{1}{1 + \frac{n_{\text{th}}}{\eta n_{\text{wav}}} }\right)^{2},
    \end{align}
    using $n_\mathrm{th}/\eta$ as a fit parameter. 
    
\begin{figure}
        \includegraphics[width = 1.0 \linewidth]{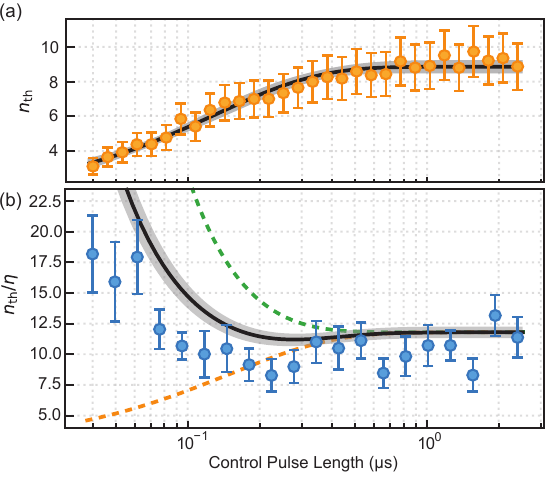} 
        \caption{\textbf{Calibration of added thermal phonons and their contribution to input-referred noise.} (a) Thermal phonon occupancy \(n_\text{th}\) of the mechanical mode as a function of control pulse duration. The black line shows an exponential saturation fit, yielding a turn-on time of \SI{150}{\nano\second}. Error bars reflect a dominant $15\%$ systematic uncertainty in the measured detection path efficiency. The gray shaded region indicates the uncertainty in the fitted model. (b) Added input-referred noise versus control pulse duration. The black line represents the total theoretically expected noise. The orange dashed line isolates the contribution from the pulse length dependence of the added thermal noise, while the green line accounts for the reduction in conversion efficiency due to decreasing temporal overlap between the control and signal pulses for short control pulse lengths (see Supplementary Information). Signal pulse and OMIT window bandwidth are, respectively, $\Gamma_\mathrm{sig}/2\pi=\SI{2.2}{\mega\hertz}$ and $\Gamma_\mathrm{eff}/2\pi=\SI{1.9}{\mega\hertz}$. All error bars represent one standard deviation calculated from the binomial statistics of count events during the HBT measurement.}\label{Fig4_control_pulse_length}
    \end{figure}

    Using HBT measurements to probe the input-referred thermal noise, we examine how the overlap between the signal bandwidth $\Gamma_\mathrm{sig}$ and the OMIT window bandwidth $\Gamma_\mathrm{eff}$ influences the conversion process. To isolate the effect of bandwidth matching, the delay following the conversion step is chosen to be much shorter than the mechanical lifetime ($T_\mathrm{storage}=\SI{150}{\nano\second}$), preventing degradation from mechanical decoherence. The control pulse peak power is fixed at $\SI{67}{\micro\watt}$ to generate an OMIT window of $\Gamma_\mathrm{eff}/2\pi=\qty{1.9}{MHz}$. Repeating the measurement in Fig.~\ref{Fig3_pulse_g2_and_bandwidth}(c) at different signal bandwidth $\Gamma_\mathrm{sig}$, we obtain the input-referred thermal noise $n_\mathrm{th}/\eta$ as a function of the signal bandwidth (see Fig.~\ref{Fig3_pulse_g2_and_bandwidth}(d)). We find a minimum of $n_\mathrm{th}/\eta \approx 12$ at $\Gamma_\mathrm{sig} \approx \Gamma_\mathrm{eff}$. For control pulses much longer than the signal pulse and in the regime of strong optical pumping $\Gamma_\mathrm{eff}\approx\Gamma_\mathrm{opt}$, the photon-phonon conversion efficiency is expected to follow (see Supplementary Information)
    \begin{align}
    \label{eq_bandwidth_matching_theory}
        \eta = \eta_{\text{c}}\frac{4\Gamma_{\text{eff}}\Gamma_{\text{sig}}}{(\Gamma_{\text{eff}} + \Gamma_{\text{sig}})^2}.
    \end{align}
    Equation~\eqref{eq_bandwidth_matching_theory} predicts a maximum efficiency of $\eta = \eta_\mathrm{c}$ when the signal bandwidth matches the OMIT bandwidth, where $\eta_\mathrm{c} = \kappa_\mathrm{e}/\kappa = 0.76$ is the external coupling efficiency of the optical cavity mode. As shown by the solid black line in Fig.~\ref{Fig3_pulse_g2_and_bandwidth}(d), the theoretical prediction from Eq.~\eqref{eq_bandwidth_matching_theory} is in good agreement with the experimental data, where we use optomechanical sideband thermometry to independently calibrate the added thermal noise $n_\mathrm{th}=9.0$ induced by the strong control pulse (see Fig.~\ref{Fig4_control_pulse_length}(a) and Supplementary Information). For an increased OMIT bandwidth of $\Gamma_\mathrm{eff}/2\pi=\SI{4.5}{\mega\hertz}$ we obtain very similar noise performance (see Supplementary Information). To ensure optimal bandwidth matching, the remaining measurements were carried out under the condition $\Gamma_\mathrm{sig} \approx \Gamma_\mathrm{eff}$.
    
    Next, we investigate the impact of thermal noise added through optical absorption during the control pulse on the input-referred noise performance of the OMIT conversion process. Longer pulses are expected to inject more energy, and thus increase the mechanical mode's thermal occupancy. Short pulses, by contrast, limit this heating and enable conversion with reduced thermal noise. We first use a pump-probe scheme to calibrate the mechanical mode occupancy as a function of control pulse duration via optomechanical sideband thermometry (see Supplementary Information). As shown in Fig.~\ref{Fig4_control_pulse_length}(a), $n_\mathrm{th}$ increases with control pulse duration and eventually saturates when absorption-driven heating is balanced by the sideband-cooling effect of the red-detuned control pulse.
    
    Based on the measurements in Fig.~\ref{Fig4_control_pulse_length}(a), one might assume that reducing the control pulse length may lead to a reduction of input-referred thermal noise in the OMIT conversion process. To investigate this hypothesis, we characterize $n_\mathrm{th}/\eta$ as a function of the control pulse duration (see Fig.~\ref{Fig4_control_pulse_length}(b)). For pulse durations longer than \SI{400}{\nano\second}, the added noise reaches a plateau, reflecting the saturation of absorption-induced mechanical thermal noise. For shorter pulses, however, the input-referred noise does not continue to decrease with decreasing $n_\mathrm{th}$; instead, it exhibits a shallow minimum before sharply increasing. This rise at short control pulse lengths is attributed to a temporal overlap mismatch between the control pulse and the signal: shortening the control pulse beyond $\sim1/\Gamma_{\text{sig}}$ reduces the fraction of the signal that is effectively mapped to the mechanical mode (see Supplementary Information). The solid black line in Fig.~\ref{Fig4_control_pulse_length}(b) shows the theoretically expected input-referred noise including the reduction of conversion efficiency due to temporal overlap mismatch and pulse-length-dependent thermal noise calibrated from Fig.~\ref{Fig4_control_pulse_length}(a). The green and orange lines isolate each of the two contributions, respectively.

\begin{figure}
        \includegraphics[width = 1.0 \linewidth]{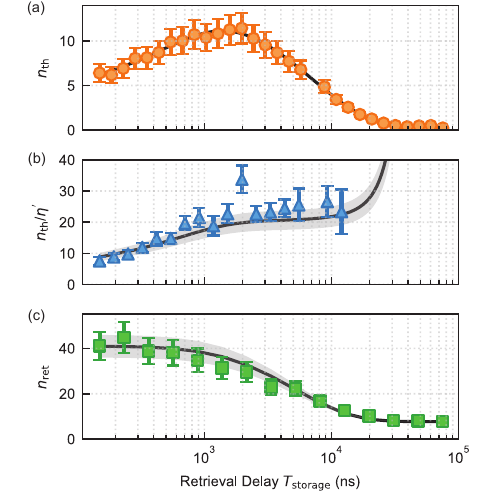} 
        \caption{\textbf{Temporal coherence of the converted mechanical excitation.} (a) Thermal phonon occupancy as a function of time delay between the $\SI{300}{ns}$-long pump pulse and the $\SI{40}{\nano\second}$-long probe pulse, where the pump pulse is the same configuration as the control pulse used in OMIT photon-phonon conversion measurements. The solid line represents the fit to a phenomenological model discussed in the text $n_\mathrm{th} = n_\mathrm{offset}-a\cdot e^{-T_\mathrm{storage}/T_\mathrm{rise}} + b\cdot e^{-T_\mathrm{storage}/T_\mathrm{1}}$ with $T_\mathrm{rise} \approx \SI{570}{ns}$ and $T_1\approx{\SI{7.3}{\mu\second}}$. The probed thermal phonon number contains the instantaneous heating from the weak probe pulse, which is negligible compared to the heating from the pump pulse. (b) Input-referred thermal noise of the OMIT optomechanical interface ($n_\mathrm{th}/\eta'$) as a function of storage time $T_\mathrm{storage}$ between the conversion pulse and the retrieval pulse, where $\eta' = \eta \cdot e^{- T_\mathrm{storage}/T_\mathrm{1}}$. The solid line represents the theoretically expected dependence based on the measured thermal phonon occupancy from (a). (c) Retrieved phonon number $n_\mathrm{ret}$ as a function of storage time shown on the left (right) y-axis for an input signal photon number $n_\mathrm{wav}= 96\pm5$. The retrieval pulse used here has nominal anti-Stokes scattering probabilities $p_\mathrm{aS} \approx 1$. The solid line represents a fit considering three contributions to the retrieved phonons as described in the text, yielding an overall conversion and retrieval efficiency of $0.35$. The error bars in (a) and (c) originate from systematic errors in the calibration of the detection path efficiency (see Supplementary Information). The error bars in (b) originate from binomial counting statistics of the HBT measurement. All error bars represent one standard deviation.}
        \label{Fig5_delay_sweep}
        \end{figure}

	As a next step, we characterize the converted mechanical state as a function of delay between the conversion and readout pulses to measure its temporal coherence properties. We first calibrate how the incurred thermal noise changes with the time delay between the conversion and retrieval pulses. Due to delayed heating, the thermal phonon number will continue to increase even after the optical pulses have stopped~\cite{wallucks_quantum_2020}. As shown in Fig.~\ref{Fig5_delay_sweep}(a), the thermal phonon occupancy of the mechanical mode is characterized by two exponential functions, where the first exponential rise indicates the time-scale of the delayed heating ($T_\mathrm{rise} \approx\SI{570}{\nano\second}$), and the second exponential decay characterizes the energy decay of the mechanical phonons, with phonon lifetime $T_\mathrm{1} \approx \SI{7.3}{\mu\second}$.
    
    Next, we proceed to characterize how the input-referred thermal noise $n_\mathrm{th}/\eta$ changes with increasing delay time between the conversion pulse and the retrieval pulse. The thermal phonon number $n_\mathrm{th}$ of the system should follow the trend in Fig.~\ref{Fig5_delay_sweep}(a), and the stored mechanical state decays exponentially, resulting in an effective conversion efficiency $\eta' = \eta \cdot e^{- T_\mathrm{storage}/T_\mathrm{1}}$. As shown in Fig.~\ref{Fig5_delay_sweep}(b), the effective input-referred thermal noise $n_\mathrm{th}/\eta'$ increases with slightly longer delays due to the delayed heating; then it reaches a plateau once the thermal phonons and stored mechanical excitation decay with the same timescale $T_\mathrm{1}$. At long delays, the input-referred thermal noise $n_\mathrm{th}/\eta'$ is no longer meaningful since most of the stored mechanical excitation has already decayed.
    
    Finally, we measure the overall photon-phonon conversion and retrieval efficiency of the optomechanical interface, by reading out the mechanical excitation via a long red-detuned retrieval pulse with nominal anti-Stokes scattering probability $p_\mathrm{aS} \approx 1$ (see Fig.~\ref{Fig5_delay_sweep}(c)). The retrieved phonon number $n_\mathrm{ret}$ is obtained via $n_\mathrm{ret} = C_\mathrm{ret}/\eta_{\text{tot}}$, where $\eta_{\text{tot}}$ is the total path efficiency and $C_\mathrm{ret}$ is the click probability of the SNSPD during the retrieval pulse (see Supplementary Information). The retrieved phonon number is comprised of three contributions: (1) coherent phonons $n_\mathrm{coh}$ originating from the converted optical input signal; (2) thermal phonons from the control pulse $n_\mathrm{pre}$, which follow the dependence in Fig.~\ref{Fig5_delay_sweep}(a); (3) instantaneous thermal phonons from the readout pulse which add an offset $n_\mathrm{0}$ to the total number of retrieved phonons. From fitting the data in Fig.~\ref{Fig5_delay_sweep}(c) using a model that includes all three contributions (see Supplementary Information), we obtain the efficiency of the retrieval process $\eta_\mathrm{r}=0.46$ and the total conversion-retrieval efficiency $\eta_\mathrm{cr} = \eta_\mathrm{r}\eta = 0.35$. This value is lower than the expected overall efficiency of $\eta_\mathrm{tot} = \eta_\mathrm{c}^2 = 0.58$, likely due to partial decay of the mapped mechanical excitation during the durations of the optical pulses as well as a reduction of the readout anti-Stokes scattering probability below the nominal value due to shifts of the optical cavity resonance frequency commonly observed in OMC devices in response to strong optical pulses~\cite{chan_laser_2011,chen_bandwidth_tunable_2024}.
\section*{Discussion}
    We have demonstrated OMIT conversion of weak coherent optical signals at the few-photon level into the mechanical mode of an OMC device. While the weak-coherent states used as input signals of the OMIT process are classical, they provide a sensitive probe of the optomechanical interface, allowing us to quantify conversion efficiency and added thermal noise and thereby assess the suitability of the device for future operation with non-classical optical states. To the best of our knowledge, our device operates at the highest conversion efficiency of $\eta=0.76$ and highest bandwidth of up to $\SI{4.5}{\mega\hertz}$ of any OMIT-based device to date. Compared to previous demonstrations of OMIT conversion with mechanical mode occupations of $n_\mathrm{th}=10^3$ to $10^6$ at room temperature, our results show a reduction by at least two orders of magnitude in thermal noise to $n_\mathrm{th}=9.0$. While we specifically choose an OMC device with short phonon lifetime of $T_\mathrm{1}=\SI{7.3}{\micro\second}$ to increase the repetition rate of the HBT measurements, two-dimensional OMC devices with lifetimes up to $\SI{20}{\milli\second}$ have been demonstrated in literature~\cite{chen_bandwidth_tunable_2024,Ren2020}. Such devices would match previous demonstrations of OMIT conversion protocols based on soft-clamped silicon nitride membranes with $T_\mathrm{1}=\SI{23}{\milli\second}$~\cite{kristensen_2024}. Notably, these previous demonstrations operate at a much narrower bandwidth of $\sim$$\SI{1.3}{kHz}$ and have only been implemented using large classical input signals with macroscopic photon numbers limited by thermal mechanical noise.
    
    The high efficiency of the OMIT conversion process enables coherent photon-to-phonon state transfer in our device, providing a route toward the preparation of non-classical mechanical states with applications in fundamental tests of quantum mechanics, quantum-enhanced metrology, and distributed quantum information protocols. In particular, OMC devices employing the OMIT conversion protocol can serve as quantum memories for optical quantum states~\cite{lvovsky_optical_2009}, enabling long-range entanglement distribution through quantum repeater protocols~\cite{duan_long-distance_2001, briegel1998quantum}. Importantly for this application, and in contrast to previous demonstrations, the large OMIT conversion bandwidth in the MHz range enabled by the strong optomechanical coupling in the OMC makes our device naturally compatible with storage of genuine single photons from telecom quantum emitters~\cite{ourari_indistinguishable_2023, yu_frequency_2023, gritsch_purcell_2023} and single-photon sources based on spontaneous four-wave mixing~\cite{chen_2024}, which exhibit comparable linewidths.
    
	The added thermal noise of the OMIT conversion process can be further reduced below the level of $n_\mathrm{th} \approx 9.0$ through several approaches: recent studies have shown a factor of six reduction in thermal noise in evanescently coupled 2D OMC devices compared to the end-coupled geometry used in this work~\cite{sonar_high-efficiency_2025}, which would lead to a projected reduction of added thermal noise in OMIT conversion to $n_\mathrm{th} \approx 1.5$. Implementing the conversion and retrieval of single-photon states in OMC devices will require even further reduction of thermal noise to the level of $n_\mathrm{th}\approx 0.3$ (see Supplementary Information). To this end, recent efforts to find materials with low optical absorption suited for low-noise optomechanical devices, such as diamond~\cite{burek_diamond_2016}, gallium phosphide~\cite{stockill_gallium_2019, schneider_optomechanics_2019, tamaki_two-dimensional_2024}, and silicon carbide~\cite{sementilli_low-dissipation_2025} are required to bring deterministic single-phonon level state preparation through OMIT-based mapping of single-photon states into reach. Overall, our results establish OMC devices as a versatile platform for the engineering of phononic quantum states and as a building block for future quantum networks, providing a coherent interface between optical photons at flexible operating wavelengths and long-lived mechanical modes suitable for quantum information storage.

\medskip

\textbf{Acknowledgments}
    We acknowledge assistance from the Kavli Nanolab Delft. This work is financially supported by the European Research Council (ERC CoG Q-ECHOS, 101001005), the Kavli Foundation (Kavli Institute Innovation Fund, QM4QT) and by the Netherlands Organization for Scientific Research (NWO) as part of the Quantum Limits grant (SUMMIT.1.016). P.V.P.N and T.P.M.A acknowledge the São Paulo Research Foundation (FAPESP) through grants 25/15127-3, 24/06827-9, 24/16573-4, 20/15786-3, 18/15580-6, 18/25339-4, and Coordenação de Aperfeiçoamento de Pessoal de Nível Superior - Brasil (CAPES) (Finance Code 001). J.N.K acknowledges the support from the European Commission for a Marie Sk\l{}odowska-Curie individual fellowship No.\ 101208412 (TPnCs\_for\_QIT).

\clearpage
\onecolumngrid
\clearpage

\setcounter{figure}{0}
\renewcommand{\thefigure}{S\arabic{figure}}
\setcounter{equation}{0}
\renewcommand{\theequation}{S\arabic{equation}}

\clearpage
\newpage

\begin{center}
	\textsc{\Large Supplementary Information} 
\end{center}
\label{SI}

\section{Device fabrication}
    \label{fabrication}
    The optomechanical crystal (OMC) devices are fabricated from silicon-on-insulator (SOI) wafers following standard silicon fabrication procedures:
    \begin{enumerate}
        \item We use SOI wafers with a silicon device layer thickness of $d_\text{Si}=\qty{250}{nm}$ and buried oxide (BOX) layer thickness of $d_\text{BOX}=\qty{3}{\micro m}$.
        \item The 2D-OMC structures are patterned in the device layer by using electron beam lithography with AR-P 6200 polymer resist and HBr/Ar reactive ion etching (RIE).
        \item After RIE, we strip the resist from the sample using dimethylformamide at $\qty{80}{\degree}$ for $\qty{10}{min}$ and a piranha solution at $\qty{70}{\degree}$ for $\qty{8}{min}$.
        \item Finally, devices are suspended by wet etching of the BOX layer using liquid hydrofluoric acid (HF, $40$~\%) for $\qty{6.5}{min}$.
        \item Before loading the samples in the dilution refrigerator, we perform another piranha cleaning and lastly remove any oxide on the silicon surface using a $\qty{1}{min}$ etch in diluted HF ($1$~\%).
    \end{enumerate}
\section{Optical measurement setup}\label{sec:SI_Optical Setup}
\begin{figure}
    	\includegraphics[width = 1.0 \linewidth]{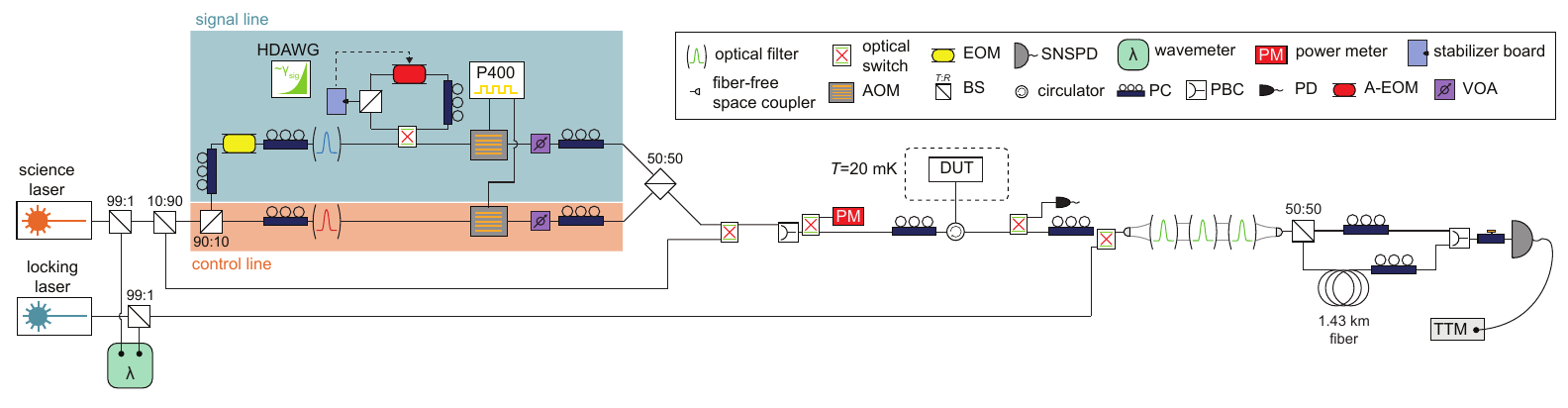} 
    	\caption{\textbf{Optical measurement setup.} Schematic of the optical setup used for pulsed OMIT measurements. BS, beam splitter with transmission (reflection) coefficient $T$ ($R$); AOM, acousto-optic modulator; EOM, electro-optic modulator; TTM, time-tagging module; PC, polarization controller; OA, optical attenuator; PM, optical power meter; PBC, polarizing beam combiner; PM, optical power meter; DUT, device under test; VOA, variable optical attenuator.}
    	\label{FigS2_optical_setup}
    \end{figure}

    Figure~\ref{FigS2_optical_setup} shows a detailed schematic of the experimental setup used for continuous-wave (cw) device characterization and pulsed optomechanically induced transparency (OMIT) measurements. Two tunable cw external diode lasers (labeled `science' and `locking' laser) are wavelength stabilized by active feedback using a wavemeter. For the pulsed OMIT measurements in the main text, the science laser directly provides the control field with frequency $\omega_\text{con}$ and is locked to the red optomechanical sideband with detuning $\Delta_\text{con} = \omega_\text{con}-\omega_\text{c}=-\Omega_\text{m}$ from the optical cavity resonance of the device at $\Omega_\text{c}$, where $\Omega_\text{m}$ is the mechanical frequency of the breathing mode. We split the light from the science laser into two optical lines for generating the control and signal pulses for the OMIT measurements using a 90:10 beam splitter. The locking laser is locked at the optical resonance frequency of the device.
    
    On the control line, the light from the science laser is filtered using a fiber-coupled Fabry--P\'erot cavity with a $\qty{50}{MHz}$ linewidth to suppress laser noise at Gigahertz frequency. We use $\qty{110}{MHz}$ acousto-optic modulators (AOMs) gated by a P400 pulse generator to generate pulses from the science laser.
    
    On the signal line, we modulate the laser using an electro-optic phase modulator (EOM) at frequency $\Omega_\text{m}$ to generate sidebands. The blue sideband generated by the EOM is at the optical cavity resonance frequency of the device. We suppress the carrier and red sideband generated by the EOM by locking a fiber-coupled Fabry--P\'erot cavity with a bandwidth of $\qty{50}{MHz}$ to the blue sideband. Through a MEMS optical switch, the signal light is routed to an amplitude-EOM, which is gated by a Zurich Instruments HDAWG arbitrary waveform generator to perform pulse shaping of the signal pulse with an exponentially rising wavepacket envelope. The bias point of the amplitude-EOM is stabilized by active feedback using a stabilizer board.
    
    The laser pulses from the signal and control line are then combined on a fiber beam splitter and sent to the OMC device under test inside the dilution refrigerator at base temperature $T=\qty{20}{mK}$ through a polarizing beam combiner (PBC) and a fiber-based circulator. Furthermore, we use an inline optical power meter and variable optical attenuators in the pulse generation path to set the pulse power before each measurement. The light coming back from the device is filtered using a series of three consecutive home-built free-space Fabry--P\'erot cavities each with linewidth $\qty{150}{MHz}$~\cite{Vlassov_Filter_Cavity_Design_2022}. The filter cavities are locked to the frequency of the optical cavity of the OMC device to filter out the strong control pulse with a suppression of $\qty{110}{dB}$ and only pass optomechanically scattered photons. Every $\qty{10}{s}$, the experiment is paused for an interval of $\qty{1}{s}$ to send continuous wave light from the locking laser to the filter cavities via a fiber-based MEMS optical switch to stabilize the locking point of the cavities. The filtered signal from the device containing the optomechanically scattered photons is detected on a superconducting nanowire single-photon detector (SNSPD) with dark count rate $\sim$$\qty{30}{Hz}$. The electronic signal from the SNSPDs is amplified and recorded by a time-tagging module (TTM).
    
    We calibrate the efficiencies throughout our optical setup, such as lensed fiber to device waveguide coupling ($\eta_\text{fc}=0.54$), efficiency of the optical detection path after the device ($\eta_\text{path}=0.08$) and the efficiency of the SNSPD ($\eta_\text{spd}=0.52$). Including the coupling efficiency from the optical cavity into the coupling waveguide $\eta_\text{c}=0.76$, the total detection efficiency of the optomechanically scattered photons is given as $\eta_\text{tot}=\eta_\text{c}\eta_\text{fc}\eta_\text{path}\eta_\text{spd}=0.017$. By performing the measurement of the setup efficiency multiple times, we estimate a systematic relative error in the calibration of the efficiency of $\Delta \eta_\text{tot}/\eta_\text{tot}=0.15$.
    
\section{Optomechanical device characterization}
    \subsection{Optical and mechanical mode characterization}
    \label{optical_mechanical_spectra}
\begin{figure}
            \centering
            \includegraphics{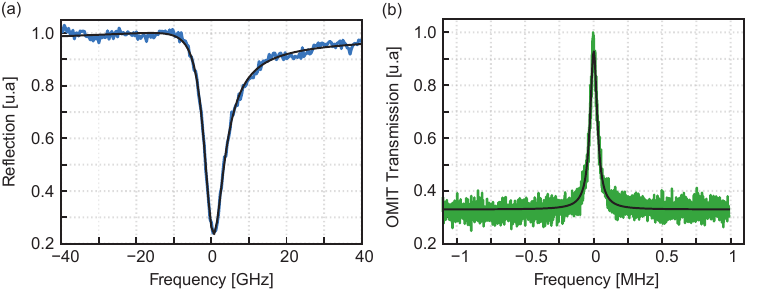}
        	\caption{\textbf{Optical and mechanical mode characterization.} (a) Normalized optical reflection spectrum obtained by scanning the laser frequency across the cavity resonance at \(\lambda_c = \SI{1530.22}{\nano\meter}\). The blue curve shows the measured response, while the black curve is a Fano-lineshape fit yielding an optical bandwidth of \(\kappa/2\pi = \SI{6.43}{\giga\hertz}\). (b) Normalized OMIT spectrum measured around the mechanical mode at \(\Omega_\text{m}/2\pi = \SI{10.314}{\giga\hertz}\). Green line indicate the measured transmission, and the solid black line is a Lorentzian fit giving a mechanical linewidth of \(\Gamma_\text{m}/2\pi = \SI{80}{\kilo\hertz}\).}
        	\label{FigS3_optical and mechanical characterization}
        \end{figure}

        Optical characterization of the device is performed by measuring the reflected power as the science laser frequency is scanned across the cavity resonance, as shown in Fig.~\ref{FigS3_optical and mechanical characterization}(a). Fiber-to-chip coupling is achieved using an end-fire coupling scheme with a lensed optical fiber. From this measurement we obtain a total optical linewidth of \(\kappa/2\pi = \SI{6.43}{\giga\hertz}\). To determine the external coupling ratio $\eta_\mathrm{c}=\kappa_\mathrm{e}/\kappa$, the laser is tuned far off-resonance and a Vector Network Analyzer (VNA) is used to drive a EOM while sweeping the generated optical sideband across the cavity. The reflected signal is detected on a high-speed photodiode connected to the VNA input, providing both the amplitude and phase response from which we extracted $\eta_\mathrm{c} = 0.76$.

        The mechanical resonance frequency $\Omega_\mathrm{m}$ and the intrinsic mechanical bandwidth $\Gamma_\mathrm{m}$ are obtained from continuous-wave OMIT measurements. The frequency driving the EOM is swept across the expected mechanical resonance, allowing us to record a high-resolution scan of the OMIT window. By operating the science laser at sufficiently low power, the optomechanical damping remains negligible, ensuring that $\Gamma_{\mathrm{eff}} \approx \Gamma_\mathrm{m}$, where $\Gamma_\text{eff}$ is the bandwidth of the OMIT window. A representative low-power OMIT trace and corresponding fit are shown in Fig.~\ref{FigS3_optical and mechanical characterization}(b) from which we extract a mechanical frequency of \(\Omega_\mathrm{m}/2\pi = \SI{10.314}{\giga\hertz}\) and an intrinsic bandwidth of \(\Gamma_\mathrm{m}/2\pi = \SI{80}{\kilo\hertz}\).
    \subsection{Calibration of single-photon optomechanical coupling $g_0$ from optomechanically scattered photon rate}
        To determine the single-photon optomechanical coupling strength $g_0$ of our device, we send laser pulses of $\qty{40}{ns}$ length on the blue optomechanical sideband to the device. We measure the count probability $C_\mathrm{S}$ of single photons from optomechanical Stokes scattering
        \begin{align}
        \label{C_stoks}
        C_\mathrm{S} = \eta_\mathrm{tot} p_\mathrm{S} (1+n_\mathrm{th}),
        \end{align}
        where $n_\mathrm{th}$ is the thermal phonon occupancy in the mechanical mode, $\eta_\mathrm{tot}$ is the total efficiency of the optical detection path and $p_\mathrm{S}$ is the intrinsic Stokes scattering probability given by
        \begin{align}
        \label{p_stoks}
        p_\mathrm{S} &= \exp \left[\frac{\kappa_\mathrm{e}}{\Omega_\mathrm{m}^2 + \kappa^2/4} \frac{4 g_0^2}{\kappa} N_\mathrm{p} \right] -1,
        \end{align}
        where $N_\mathrm{p}$ is the number of photons in the excitation pulse that enter the coupling waveguide. For low optical pump power, the mechanical mode of the device is well-thermalized to the Millikelvin environment of the dilution refrigerator and is thus close to its quantum ground state, such that we can assume $n_\mathrm{th} \approx 0$ in Eq.~\eqref{C_stoks}. Moreover, at low optical powers, we can expand the exponential in Eq.~\eqref{p_stoks} to first order. Combining these two approximations and solving for $g_0$, we can calculate $g_0$ from the measured single-photon count rate $C_\mathrm{S}$ as
        \begin{align}
        \label{g0}
            g_0 = \left( \frac{C_\mathrm{S}}
            {\eta_\mathrm{tot}}
            \frac{\Omega_\mathrm{m}^2 + \kappa^2/4}{4 N_\mathrm{p}} \frac{\kappa}{\kappa_\mathrm{e}} \right)^{1/2}.
        \end{align}
        
        The value of $g_0/2\pi=\SI{0.97}{\mega\hertz}$ is in good agreement with the simulated value $g_0/2\pi=\SI{1.0}{\mega\hertz}$ and the value obtained from the power dependence of the OMIT window bandwidth $g_0/2\pi=\SI{1.1}{\mega\hertz}$ as discussed in the main text.
    \subsection{Sideband thermometry}
    \label{sideband_thermometry}
        The mechanical phonon occupancy is extracted using single-sideband thermometry. The device is probed with pulses tuned to the red sideband of the optical cavity, $\Delta = -\Omega_\text{m}$, such that the dominant scattering process is the anti-Stokes conversion of phonons into up-shifted photons. The rate of detected optomechanically scattered photons, $C_\mathrm{aS}$, is measured using SNSPDs. The anti-Stokes scattering probability $p_\mathrm{aS}$ is set by the intracavity photon number $n_\mathrm{c}$ and the optomechanical coupling $g_0$. The phonon occupancy follows directly from $n_\mathrm{th} = C_\mathrm{aS}/(\eta_\mathrm{tot}p_\mathrm{aS})$, where $\eta_\mathrm{tot}$ is the total detection efficiency of the optical path. The scattering probability for a probe pulse containing $N_\mathrm{p}^\mathrm{(probe)}$ photons is given by
        \begin{equation}
        \label{eq_sideband_thermometry}
            p_{a\mathrm{S}} = 1 - \exp\!\left[-\,\frac{\kappa_{\mathrm{e}}}{\Omega_{\text{m}}^{2} + \kappa^{2}/4}\frac{4g_{0}^{2}}{ \kappa}N_\mathrm{p}^\mathrm{(probe)}\right].
        \end{equation}
        In the main text, we use single-sideband thermometry in a pump-probe scheme to calibrate the added thermal noise induced by the strong control pulse in the OMIT memory. The pump pulse is set to the same length $T_\mathrm{con}$ and intracavity photon number $n_\mathrm{c}$ as the control pulse and the delay between pump and probe pulse is set to the delay of the control and retrieval pulses in the OMIT measurement $T_\mathrm{storage}$. After the pump pulse, a weak probe pulse with fixed intracavity photon number $n_\mathrm{c}^{(\mathrm{probe})} = \kappa_\mathrm{e} N_\mathrm{p}^\mathrm{(probe)}/\left[(\Omega_\mathrm{m}^2+\kappa^2/4)T_\mathrm{probe}\right] \ll n_\mathrm{c}$ and pulse length $T_\mathrm{probe}=\SI{40}{\nano\second}$, probes the thermal phonon occupancy of the mechanical mode. We use the count rate of optomechanically scattered photons from the probe pulse to calculate $n_\mathrm{th}$ via $n_\mathrm{th} = C_\mathrm{aS}/(\eta_\mathrm{tot}p_\mathrm{aS})$. The probed thermal phonon number contains the instantaneous heating from the weak probe pulse, which is negligible compared to the heating from the pump pulse.
    \subsection{Mechanical lifetime measurement}
        To characterize the mechanical mode lifetime we directly measure the lifetime of a large, intentionally created thermal phonon population in the target mechanical mode. We first send a $\SI{300}{\nano \second}$ square laser pulse that is red detuned from the optical cavity resonance to the device. This pulse is purely intended to inject optical power into the cavity, which is partially absorbed in the silicon slab. The resulting local heating gives rise to a hot, non-equilibrium thermal bath that strongly couples to the mechanical mode and rapidly populates it with a large number of thermal phonons. After a variable delay time, we perform a readout of the mechanical population using a second laser pulse of duration $\SI{40}{\nano \second}$, tuned to the red optomechanical sideband of the cavity. During this pulse, anti-Stokes scattering converts phonons into upshifted photons which are collected at the output and detected using our SNSPD. The anti-Stokes scattering rate is proportional to the instantaneous phonon number and photon counting of the readout pulse provides a direct measurement of the mechanical occupation at that specific delay time. By scanning the delay between the heating and readout pulses, we reconstruct the full temporal evolution of the phonon population. 
        
        The resulting trace displayed in Fig.5(a) of the main text shows two distinct regimes. Immediately after the heating pulse, we observe a fast rise in the detected phonon population, originating from delayed heating processes that continue to feed energy into the mechanical mode for a short time after the pulse ends. After this transient heating period, the phonon population enters a clear exponential decay regime governed by the intrinsic mechanical damping. Fitting of the data yields a mechanical energy lifetime of $T_1 = \SI{7.3}{\micro\second}$.
\section{Hanbury-Brown Twiss measurements of the converted phonon state}
\begin{figure}
        \centering
        \includegraphics[width = 0.5 \linewidth]{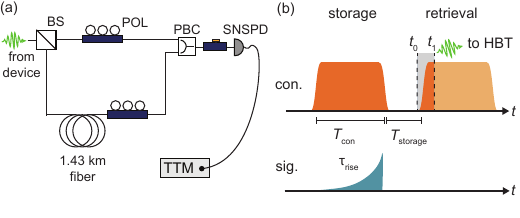} 
        \caption{\textbf{Experimental implementation of Hanbury Brown-Twiss measurement of the converted phonon state.} Optical setup for Hanbury Brown-Twiss interferometry measurements. BS, 50:50 beam splitter; POL, fiber polarization paddle; PBC, polarizing beam combiner; SNSPD, superconducting nanowire single photon detector; TTM, time-tagging module. (b) Pulse sequence of control (top) and signal pulses (bottom).}
        \label{FigS5_HBT_details}
    \end{figure}

    In the main text, we perform Hanbury Brown-Twiss (HBT) measurements to measure the second-order intensity autocorrelation function $g^{(2)}(0)$ of the mechanical mode after conversion of a weak coherent input signal. To probe $g^{(2)}(\tau)$ of the phonon state, we use the optomechanical anti-Stokes process induced by the red-detuned retrieval pulse to swap part of the mechanical state onto the optical mode and subsequently measure $g^{(2)}(\tau)$ using a fiber optical HBT setup~\cite{hong_hanbury_2017}. Our fiber optical setup for performing HBT measurements is shown in Fig.~\ref{FigS5_HBT_details}(a). At the time when the experiments presented in this work were performed, only one SNSPD was available in our lab. Therefore, we use a delay line after the 50:50 beam splitter and a polarizing beam combiner to perform HBT measurements instead of the usual setup consisting of a single 50:50 beam splitter and two SNSPDs. The $\SI{1.43}{\kilo\meter}$ long delay line induces a delay of \SI{7.146}{\micro\second} on one output port of the 50:50 beam splitter. The two output ports are combined on the polarizing beam combiner with orthogonal polarizations and then detected on the SNSPD. The SNSPD records detection events in two separate time bins D1 and D2, which act as the two detection channels of the HBT measurement.
    
    The autocorrelation function of the mechanical mode with annihilation operator $\hat{b}$ is formally defined as
    \begin{align}
    \label{g_2_definition}
        g^{(2)}(\tau) = \frac{\langle  b^{\dag}(0) b^{\dag}(\tau) b(\tau) b(0)  \rangle }
        {\langle b^{\dag}(\tau) b(\tau) \rangle
        \langle b^{\dag}(0) b(0) \rangle},
    \end{align}
     where $\tau$ is the time delay between a coincidence detection on detector channels D1 and D2 in the HBT setup. In the regime of low photon detection probability $p_\mathrm{click}$ and for $\tau=0$, $g^{(2)}(\tau)$ can be experimentally measured through the probabilities of coincidences on detector channels D1 and D2~\cite{stevens_third-order_2014,hong_hanbury_2017}:
    \begin{align}
    \label{g_2_measure}
        g^{(2)}(0) = \frac{P(\mathrm{D_1} \cap \mathrm{D_2})}{P(\mathrm{D_1}) P(\mathrm{D_2})}, 
    \end{align}
    where $P(\mathrm{D_1} \cap \mathrm{D_2})$ describes the probability of a coincidence detection on both detector channels and $P(\mathrm{D_1})$ ($P(\mathrm{D_2})$) represents the probability of a single detection event on detector channel D1 (D2).
    
    Importantly, in the measurements presented in the main text, we typically sweep the number of coherent signal photons coupled into the device waveguide $n_\mathrm{wav}$. In these measurements, the single-click probabilities $P(\mathrm{D_1})$ and $P(\mathrm{D_2})$ depend on the size of the converted weak coherent state and will thus increase with increasing $n_\mathrm{wav}$ if a fixed time interval is used to analyze the retrieved photons. However, to reliably calculate $g^{(2)}(0)$ from the measured click probabilities via Eq.~\ref{g_2_measure}, it is crucial to ensure $P(\mathrm{D_1}), P(\mathrm{D_2})\ll1$. To ensure this condition, all HBT measurements presented in the main text are performed and analyzed in the following way: to retrieve the converted mechanical state, we always send long ($T_\mathrm{ret}=\SI{2.4}{\micro\second}$) red-detuned optical pulses with the same intracavity photon number $n_\mathrm{c}=3,200$ as the control pulse. To limit the click probability, in postprocessing we only analyze the first few tens of nanoseconds of the retrieval pulse in between times $t_0$ and $t_1$ to calculate $g^{(2)}(0)$ as shown in Fig.~\ref{FigS5_HBT_details}(b). When varying $n_\mathrm{wav}$, we vary the end point of the time window used for data analysis $t_1$, such that we always have single-click probabilities $P(\mathrm{D_1})\approx P(\mathrm{D_2})\approx 0.04$. This value is chosen to ensure at the same time $P(\mathrm{D_1}),P(\mathrm{D_2})\ll 1$ while still measuring a sufficient amount of coincidence counts ($N_\mathrm{coinc} \approx 100 \text{--} 1,000$). We note that the retrieval pulse, in principle, also adds thermal phonons to the mechanical mode due to optical absorption heating. However, as the analysis time window used to calculate $g^{(2)}(0)$ from the retrieval pulse is much shorter than the control pulse ($T_\mathrm{con}=300 \text{--} \SI{2400}{\nano\second}$ in all measurements discussed in the main text), the thermal phonon noise added to the converted mechanical state is largely dominated by optical absorption heating from the control pulse, which we calibrate as described in Sec.~\ref{sideband_thermometry} from optomechanical single-sideband thermometry.
    
    To calibrate $n_\mathrm{wav}$, after each HBT measurement sequence, we switch off the control pulse and only send the signal pulse to the device. Based on the calibration of the optical path efficiencies and the reflectivity of the optical cavity for the signal pulse on cavity resonance $R_\mathrm{c} = (1-2\eta_\mathrm{c})^2$, we can calculate $n_\mathrm{wav}$ from the count probability of single-photons on the SNSPD $C_\mathrm{sig}$ as
    \begin{align}
        n_\mathrm{wav} = \frac{C_\mathrm{sig}}{R_\text{c}\eta_\text{fc}\eta_\text{path}\eta_\text{spd}}.
    \end{align}
\section{Theory of OMIT photon-phonon conversion}
    \subsection{OMIT conversion efficiency}
        A convenient starting point for describing OMIT-based conversion is the input–output formalism together with the Langevin equations for an optomechanical system with dispersive coupling~\cite{GardinerCollett1985,aspelmeyer_cavity_2014}. In the frame of the science laser, the optical and mechanical mode operators evolve as
        \begin{align}
            &\dot{a}(t) = (i\Delta - \kappa/2)a(t) + i g_{\omega}a(t)(b(t)+b^{\dagger}(t)) + \sqrt{\eta_{\text{c}}\kappa}\,a^{(e)}_{\text{in}}(t)+ \sqrt{(1-\eta_{\text{c}})\kappa}\,a^{(i)}_{\text{in}}(t),\\
            &\dot{b}(t) = (-i\Omega_{\text{m}} - \Gamma_{\text{m}}/2)b(t) + ig_{\omega}a^{\dagger}(t)a(t) + \sqrt{\Gamma_\text{m}}b^{(i)}_{\text{in}}(t),\\
            &a_{\text{out}}(t) = a^{(e)}_{\text{in}}(t) - \sqrt{\eta_\text{c}\kappa}\,a(t),
        \end{align}
        where $g_{\omega}$ is the vacuum dispersive optomechanical coupling rate and $\Delta = \omega_{\text{con}} - \omega_{\text{c}}$ is the control-laser detuning. The parameter $\eta_{\text{c}} = \kappa_{\text{e}}/\kappa$ denotes the external coupling fraction, with $\kappa_{\text{e}}$ the coupling rate to the waveguide and $\kappa$ the total optical energy decay rate, including both radiative and absorptive intrinsic losses. According to the fluctuation–dissipation theorem~\cite{Kubo1966}, each decay channel introduces a corresponding noise drive, described by the external and intrinsic input fields $a_{\text{in}}^{(e)}(t)$ and $a_{\text{in}}^{(i)}(t)$. Similarly, the mechanical mode decays at a rate $\Gamma_{\text{m}}$ and is driven by a thermal noise input $b_{\text{in}}^{(i)}(t)$. Lastly, the cavity output through the external port is given by $a_{\text{out}}(t)$. 

        Following the experimental procedure described in Sec.~\ref{sec:SI_Optical Setup}, the external drive can be modeled as $a^{(e)}_{\text{in}}(t) = \bar{a}^{\text{(e)}}_{\text{in}} + \delta a^{(e)}_{\text{in}}(t)\,e^{-i\Omega_{\text{sig}}t}$ where the first term represents the control pulse and the second term corresponds to the signal field. In the experiment, the signal field is produced by phase modulating a portion of the science laser at $\Omega_{\text{sig}}$ using an electro-optic modulator, which creates sidebands at $\pm \Omega_{\text{sig}}$. A fiber-coupled Fabry–Pérot cavity is locked to the blue sideband to suppress both the carrier and the red sideband, yielding the weak signal pulse used in the OMIT conversion protocol. By coherently driving the system, an ansatz on the optical and mechanical mode operators can be made as 
        \begin{align}
            &a(t)\xrightarrow[]{}\bar{a} + \delta a(t)e^{-i\Omega_{\text{sig}}t},\\
            &b(t)\xrightarrow[]{}\bar{b} + \delta b(t)e^{-i\Omega_{\text{sig}}t},\\
            &a_\text{out}(t)\xrightarrow[]{}\bar{a}_\text{out} + \delta a_{\text{out}}(t)e^{-i\Omega_{\text{sig}}t},
        \end{align}
        and a set of equations for the slowly varying envelopes is obtained by retaining only the terms oscillating at the signal frequency and neglecting noise contributions, justified by the fact that the system is driven by bright coherent fields:
        \begin{align}
            &\delta \dot{a}(t) = \left[i(\bar{\Delta}+\Omega_{\text{sig}})-\kappa/2\right]\delta a(t) + ig_{\omega}\bar{a}\,\delta b(t) + \sqrt{\eta_\text{c}\kappa}\,\delta a_{\text{sig}}(t),\label{Eq. dadt}\\
            &\delta \dot{b}(t) = \left[-i(\Omega_{\text{m}}-\Omega_{\text{sig}})-\Gamma_\text{m}/2\right]\delta b(t) + ig_{\omega}\bar{a}^{\dagger}\delta a(t),\label{Eq. dbdt}\\
            &\delta a_{\text{out}}(t) = \delta a_{\text{sig}}(t)-\sqrt{\eta_{\text{c}}\kappa}\,\delta a(t).\label{Eq. a_out}
        \end{align}
        Here, $\bar{\Delta} =\Delta + g_{\omega}(\bar{b}+\bar{b}^\dagger)$ denotes the renormalized detuning,, which incorporates the shift induced by the steady-state mechanical displacement, where $\bar{b} = ig_{\omega}|\bar{a}(t)|^2/(\Gamma_{\text{m}}/2+i\Omega_{\text{m}})$, and results from the optical steady-state intracavity field $\bar{a} = \sqrt{\eta_\text{c}\kappa}\,\bar{a}^{(e)}_{\text{in}}/(\kappa/2 - i\bar{\Delta})$. Formal integration of Eq.~\ref{Eq. dadt} yields
        \begin{align}
            \delta a(t) &= ig_{\omega}\bar{a}\int_{-\infty}^{\infty}\chi_{\text{c}}(t-t^\prime)\delta b(t^{\prime})\,dt^{\prime} + \sqrt{\eta_{\text{c}}\kappa}\int_{-\infty}^{\infty}\chi_{\text{c}}(t-t^\prime)\delta a_{\text{sig}}(t^\prime)\,dt^{\prime}\label{Eq. a(t)_1}\\
            &\approx i\sqrt{\frac{\Gamma_\text{opt}}{\kappa}}\,\delta b(t) + 2\sqrt{\frac{\eta_{\text{c}}}{\kappa}}\,\delta a_{\text{sig}}(t),\label{Eq. a(t)_2}
        \end{align}
        where the rotating frame form of the cavity response is defined as $\mathcal{F}\{\chi_{\text{c}}(t)\}(\omega) = 1/(\kappa/2 - i(\bar{\Delta} + \Omega_{\text{sig}} + \omega))$. By choosing $\bar{\Delta} = -\Omega_\text{m}$, the optomechanically induced damping is identified as $\Gamma_{\text{opt}} = \frac{4g_{\omega}^2|\bar{a}|^2}{\kappa}$. For a narrowband mechanical oscillator, as considered in this work, the convolution appearing in the first term of Eq.~\ref{Eq. a(t)_1} can be approximated by $\mathcal{F}\{\chi_{\text{c}}(t)\}(\Omega_{\text{m}} - \Omega_{\text{sig}})\,\delta b(t)$, since $\delta b(t)$ varies only slowly on the timescale $1/\kappa$. When $\kappa$ is large, the function $\chi_{\text{c}}(t)$ becomes sharply peaked and effectively behaves as a delta distribution when acting on functions that evolve slowly compared to $1/\kappa$. In particular, one finds that $\lim_{\kappa\to\infty} \frac{\kappa}{2}\chi_{\mathrm{c}}(t - t^\prime)= e^{-i(\bar{\Delta} + \Omega_{\mathrm{sig}})(t - t^\prime)}\,\delta(t - t^\prime)$, which justifies the approximation used in obtaining the second term of Eq.~\ref{Eq. a(t)_2}. This result allows us to express Eq.~\ref{Eq. dbdt} directly in terms of the mechanical operator and the applied signal field:
        \begin{equation}
            \delta \dot{b}(t) = \left(-i\delta - \Gamma_{\text{eff}}/2\right)\delta b(t) + i\sqrt{\Gamma_{\text{opt}}\,\eta_{\text{c}}}\,\delta a_{\text{sig}}(t),\label{Eq. dbdt_2}
        \end{equation}
        where $\delta = \Omega_\mathrm{m} - \Omega_{\text{sig}}$ is the two-photon detuning and $\Gamma_{\text{eff}} = \Gamma_{\text{m}} +\Gamma_{\text{opt}}$ is the effective mechanical oscillator bandwidth.

        To calculate the conversion efficiency, the mechanical evolution must be tracked up to the retrieval stage. Following the procedure in~\cite{kristensen_2024}, we initially consider the dynamics for times $t < 0$, during which both the signal and control fields are present. In this work, we also include the case where the signal and control pulses are temporally mismatched as in the case of when the control pulse is much shorter than the signal. We assume that the pulses are aligned such that both fields are switched off simultaneously at t = 0, with the control pulse having duration $t_0$ and that $\delta b(t_0) = 0$. Under these conditions, formal integration of Eq.~\ref{Eq. dbdt_2} yields
        \begin{equation}
            \delta b(t<0) = i\sqrt{\Gamma_{\text{opt}}\eta_{\text{c}}}\,e^{-(i\delta + \Gamma_{\text{eff}}/2)t}\int_{t_0}^{t}d\tau \,e^{(i\delta + \Gamma_{\text{eff}}/2)\tau} \delta a_{\text{sig}}(\tau),
        \end{equation}
        where, for an exponentially rising signal field with time constant $2/\Gamma_\text{sig}$, $\delta a_{\text{sig}}(t) = \delta a_{\text{sig}}(0) \Theta(-t)e^{\Gamma_{\text{sig}}t/2}$, the mechanical amplitude at $t = 0$ is given by
        \begin{equation}
            \delta b(0) = \frac{i\sqrt{\Gamma_{\text{opt}}\eta_{\text{c}}}}{\frac{\Gamma_{\text{eff}} + \Gamma_{\text{sig}}}{2}+i\delta}\left[1-\exp{\left(\left(i\delta+\frac{\Gamma_{\text{eff}} + \Gamma_{\text{sig}}}{2}\right)t_0\right)}\right]\delta a_{\text{sig}}(0).
        \end{equation}
    
        After the mapping stage, both the signal and control fields are switched off, leaving the mechanical excitation to evolve freely. During the storage interval $T_{\text{storage}}$, the mechanical amplitude decays solely under its intrinsic lifetime, $\delta b(T_{\text{storage}}) = \delta b(0)e^{-i\delta T_\text{storage}}e^{-\Gamma_\text{m}T_\text{storage}/2}$. Retrieval is initiated by reapplying the control field, which activates the anti-Stokes process. Once the control is on, the mechanical mode experiences an optomechanically enhanced damping rate, leading to, $\delta b(t > T_{\text{storage}}) = \delta b(T_\text{storage})e^{-(i\delta+\Gamma_\text{eff}/2)(t-T_{\text{storage}})}$, as the mapped phonons are converted back into optical photons. The output field during retrieval is obtained by combining Eqs.~\ref{Eq. a_out} and \ref{Eq. a(t)_2}, taking into account that the signal field is absent in this stage, $\delta a_{\mathrm{sig}}(t) = 0$:
        \begin{align}
            \delta a_{\text{out}}(t>T_{\text{storage}})&= -i\sqrt{\eta_{\text{c}}\kappa}\sqrt{\frac{\Gamma_\text{opt}}{\kappa}}\,\delta b(t>T_\text{storage})=\\ &=\frac{\eta_\text{c}\Gamma_{\text{opt}}\delta a_{\text{sig}}(0)}{\frac{\Gamma_{\text{eff}} + \Gamma_{\text{sig}}}{2}+i\delta}\left[1-\exp{\left(\left(i\delta+\frac{\Gamma_{\text{eff}} + \Gamma_{\text{sig}}}{2}\right)t_0\right)}\right] e^{-i\delta T_\text{storage}}e^{-\Gamma_\text{m}T_\text{storage}/2} e^{-(i\delta+\Gamma_\text{eff}/2)(t-T_{\text{storage}})}.\label{Eq. aout}
        \end{align}
        
        For sufficiently long retrieval pulses, the anti-Stokes process converts the entire mapped mechanical excitation into optical photons at the cavity resonance. These photons exit through the waveguide with efficiency $\eta_\text{c}$ so the total energy collected at the output is $E_\text{out} = \eta_{\text{c}}E_{\text{str}}$ where $E_{\text{str}}$ denotes the converted energy. The conversion efficiency is therefore $\eta \equiv E_{\text{str}}/E_{\text{in}}= (1/\eta_\mathrm{c})E_{\text{out}}/E_{\text{in}}$, where $E_{\text{in}}$ denotes the total input energy contained in the signal pulse. The total input energy and output energy is, respectively,
        \begin{align}
            E_{\text{in}} &= \int_{-\infty}^{0}|\delta a_{\text{sig}}(t)|^2 dt = \frac{|\delta a_{\text{sig}}(0)|^2}{\Gamma_{\text{sig}}}\\
            E_{\text{out}} &= \int_{T_{\text{storage}}}^{\infty}|\delta a_{\text{out}}(t)|^2 dt =\\
            &=|\delta a_{\text{sig}}(0)|^2 e^{-\Gamma_\text{m}T_\text{storage}}\frac{4\eta_{\text{c}}^2\Gamma_{\text{opt}}^2/\Gamma_{\text{eff}}}{(\Gamma_{\text{eff}} + \Gamma_{\text{sig}})^2 + 4\delta^2}\left|1-\exp{\left(\left(i\delta+\frac{\Gamma_{\text{eff}} + \Gamma_{\text{sig}}}{2}\right)t_0\right)}\right|^2.
            \label{temporal-mode overlapping factor}
        \end{align}
        Finally, in the regime relevant to this work, where $\Gamma_{\text{m}} \ll \Gamma_{\text{opt}}$ and so $\Gamma_\mathrm{eff}\approx\Gamma_\mathrm{opt}$ and the signal is tuned to two-photon resonance ($\delta$ = 0), the conversion efficiency reduces to
        \begin{align}
            \eta = \eta_{\text{c}}\frac{4\Gamma_{\text{eff}}\Gamma_{\text{sig}}}{(\Gamma_{\text{eff}} + \Gamma_{\text{sig}})^2}\left|1-\exp{\left(\left(\frac{\Gamma_{\text{eff}} + \Gamma_{\text{sig}}}{2}\right)t_0\right)}\right|^2e^{-\Gamma_\text{m}T_\text{storage}}.\label{Eq. eta_storage}
        \end{align}
        
        For sufficiently long control pulses, Eq.\ref{Eq. eta_storage} reduces to the form reported in~\cite{kristensen_2024}. In this regime, the temporal-mode overlap between the control and signal fields approaches unity, causing the term inside the modulus to converge to one. Moreover, in most measurements presented in this work, the delay between the conversion and retrieval pulses is much shorter than the mechanical lifetime, so the exponential decay factor is also effectively unity. Under these two conditions, Eq.~\ref{Eq. eta_storage} simplifies to the following expression for the conversion efficiency:
        \begin{equation}
            \eta = \eta_{\text{c}}\frac{4\Gamma_{\text{eff}}\Gamma_{\text{sig}}}{(\Gamma_{\text{eff}} + \Gamma_{\text{sig}})^2}.\label{Eq. eta_storage_simplfied}
        \end{equation}
    \subsection{Autocorrelation function of retrieved state}
        In this section, we derive expressions for the measured autocorrelation function $g^{(2)}(0)$ of a displaced thermal state as considered in the measurements in the main text. Furthermore, we investigate the case of a single-phonon Fock state mapped in the mechanical mode with added thermal noise to derive a criterion for realizing single-photon conversion in an OMIT-based photon-phonon interface.
        \subsubsection{Displaced thermal states}
            In the measurements presented in the main text, we perform an HBT measurement to measure the second-order intensity autocorrelation function $g^{(2)}(\tau)$ of the phonon mode, which is in a displaced thermal state after the conversion of the weak coherent input signal. We derive the expected $g^{(2)}(0)$ for the displaced thermal state based on a derivation in~\cite{forsch_microwave--optics_2020}. The mechanical field $\hat{b}$ is composed of a coherent component $\beta$ and a thermal fluctuation represented by a zero-mean Gaussian field $\hat{\delta}(\tau)$.
            \begin{align}
                \label{coherent_storage_mechanics}
                \hat{b} = \beta + \hat{\delta}(\tau),
            \end{align}
            where $|\beta|^2=n_\text{coh}$ and $n_\text{coh}$ is the number of coherent phonons transferred to the mechanical mode in the OMIT conversion process related to the input signal photon number coupled into the device waveguide $n_\text{wav}$ through the conversion efficiency $\eta$ as $n_\text{coh}=\eta n_\text{wav}$. The zero-mean thermal field has the properties
            \begin{align}
                \langle \hat{\delta}(t) \rangle &= 0, \label{zero_mean}\\[4pt]
                G(\tau) &\equiv \langle \hat{\delta}^\dagger(t)\,\hat{\delta}(t+\tau) \rangle 
                = n_{\text{th}}\, g^{(1)}_{\text{th}}(\tau), \\[4pt]
                \langle \hat{\delta}(t)\,\hat{\delta}(t') \rangle 
                &= \langle \hat{\delta}^\dagger(t)\,\hat{\delta}^\dagger(t') \rangle = 0,
            \end{align}
            with $g^{(1)}(\tau)$ the first-order field correlation function of the thermal state and average thermal phonon occupancy $n_\text{th}$ corresponding to the thermal noise added in the OMIT conversion process.
            
            We calculate the autocorrelation function of the mechanical mode defined as
            \begin{align}
                \label{g2_definition}
                g^{(2)}(\tau)
                &= \frac{%
                \left\langle
                \hat{b}^\dagger(t)\,
                \hat{b}^\dagger(t+\tau)\,
                \hat{b}(t+\tau)\,
                \hat{b}(t)
                \right\rangle%
                }{%
                \left\langle
                \hat{b}^\dagger(t)\,\hat{b}(t)
                \right\rangle^{2}
                }.
            \end{align}
            Inserting Eq.~\eqref{coherent_storage_mechanics} into Eq.~\eqref{g2_definition}, we explicitly calculate the Gaussian moments in the numerator and denominator of Eq.~\eqref{g2_definition}. For the numerator, we obtain
            \begin{align}
                \left\langle
                \hat{b}^\dagger(t)\,
                \hat{b}^\dagger(t')\,
                \hat{b}(t')\,
                \hat{b}(t)
                \right\rangle
                &=
                \left\langle
                (\beta^* + \hat{\delta}^\dagger(t))\,
                (\beta^* + \hat{\delta}^\dagger(t'))\,
                (\beta + \hat{\delta}(t'))\,
                (\beta + \hat{\delta}(t))
                \right\rangle
                \label{eq:S6}
                \\
                &=
                \begin{aligned}[t]
                    &|\beta|^4 \\[4pt]
                    &\quad
                    + |\beta|^2 \Big(
                    \langle \hat{\delta}^\dagger(t)\hat{\delta}(t)\rangle
                    + \langle \hat{\delta}^\dagger(t')\hat{\delta}(t')\rangle
                    + \langle \hat{\delta}^\dagger(t)\hat{\delta}(t')\rangle
                    + \langle \hat{\delta}^\dagger(t')\hat{\delta}(t)\rangle
                    \Big) \\[4pt]
                    &\quad
                    + \langle
                    \hat{\delta}^\dagger(t)\,
                    \hat{\delta}^\dagger(t')\,
                    \hat{\delta}(t')\,
                    \hat{\delta}(t)
                    \rangle
                \end{aligned}
                \label{eq:S7}
            \end{align}
            with $t' = t + \tau$. We note that in Eq.~\eqref{eq:S7} moments with odd orders of $\hat{\delta}$ vanish as the thermal field has zero mean (see Eq.~\eqref{zero_mean}). The second term in Eq.\eqref{eq:S7} simplifies to
            \begin{align}
                \label{denominator_2}
                |\beta|^{2} \Big(
                \underbrace{
                \langle \hat{\delta}^\dagger(t)\hat{\delta}(t)\rangle
                + \langle \hat{\delta}^\dagger(t')\hat{\delta}(t')\rangle
                }_{=\,2n_{\text{th}}}
                +\,
                \underbrace{
                \langle \hat{\delta}^\dagger(t)\hat{\delta}(t')\rangle
                + \langle \hat{\delta}^\dagger(t')\hat{\delta}(t)\rangle
                }_{=\,G(\tau)+G^*(\tau)}
                \Big)
                &=
                2|\beta|^{2}n_{\text{th}}
                + |\beta|^{2}\big[G(\tau) + G^{*}(\tau)\big].
            \end{align}
            Note that the term proportional to $\big[G(\tau) + G^{*}(\tau)\big]$ is phase-sensitive and will thus drop out in photodetection when performing an ensemble average over the relative phases in the thermal state. Using Gaussian moment factoring on the third term in Eq.~\eqref{eq:S7}, we obtain
            \begin{align}
                \left\langle
                \hat{\delta}^\dagger(t)\,
                \hat{\delta}^\dagger(t')\,
                \hat{\delta}(t')\,
                \hat{\delta}(t)
                \right\rangle
                &=
                \langle \hat{\delta}^\dagger(t)\hat{\delta}(t)\rangle
                \langle \hat{\delta}^\dagger(t')\hat{\delta}(t')\rangle
                +
                \langle \hat{\delta}^\dagger(t')\hat{\delta}(t)\rangle
                \langle \hat{\delta}^\dagger(t)\hat{\delta}(t')\rangle, \\[6pt]
                &=
                n_{\text{th}}^{2}
                + |G(\tau)|^{2}, \\[6pt]
                &= n_{\text{th}}^{2}
                \!\left[1 + \big|g_{\text{th}}^{(1)}(\tau)\big|^{2}\right].\label{denominator_3}
            \end{align}
            For the denominator in Eq.~\eqref{eq:S7} we find
            \begin{align}
                \langle \hat{b}^\dagger \hat{b} \rangle
                &=
                \left\langle
                (\beta^* + \hat{\delta}^\dagger)
                (\beta + \hat{\delta})
                \right\rangle \\[4pt]
                &=
                |\beta|^{2} + n_{\text{th}} \\[6pt]
                \Rightarrow\quad
                \langle \hat{b}^\dagger \hat{b} \rangle^{2}
                &=
                \left(n_{\text{coh}} + n_{\text{th}}\right)^{2}.\label{numerator}
            \end{align}
            Inserting the results from Eqs.~\eqref{eq:S7},\eqref{denominator_2}, and~\eqref{denominator_3} into Eq.~\eqref{g2_definition}, we find the full expression of the autocorrelation function of a displaced thermal state as
            \begin{align}
                g^{(2)}(\tau)&=1+ \frac{n_{\text{th}}^{2} + 2 n_{\text{coh}}n_{\text{th}}}{\left( n_{\text{coh}} + n_{\text{th}}\right)^{2}}\,\big| g_{\text{th}}^{(1)}(\tau) \big|^{2}\\[6pt]
                &=1 +\left(1 -\left(\frac{ n_{\text{coh}} }{ n_{\text{coh}} + n_{\text{th}}}\right)^{2}\right)\big| g_{\text{th}}^{(1)}(\tau)\big|^{2}.
            \end{align}
            At $\tau=0$, $|g_{\text{th}}^{(1)}(0)| = 1$ so we find
            \begin{align}
                g^{(2)}(0)
                &=
                2 -
                \left(
                \frac{ n_{\text{coh}} }{ n_{\text{coh}} + n_{\text{th}} }
                \right)^{2}\\
                &=
                2 -
                \left(
                \frac{1}{1 + \frac{n_{\text{th}}}{n_{\text{coh}}} }
                \right)^{2}\\
                &=
                2 -
                \left(
                \frac{1}{1 + \frac{n_{\text{th}}}{\eta n_{\text{wav}}} }
                \right)^{2}.\label{g2_displaced_thermal}
            \end{align}
            Equation~\eqref{g2_displaced_thermal} is used to fit the data for the HBT measurement using $n_\text{th}/\eta$ as the fitting parameter.
        \subsubsection{Single photon conversion in the presence of thermal noise}
\begin{figure}
            	\includegraphics[width = 0.5 \linewidth]{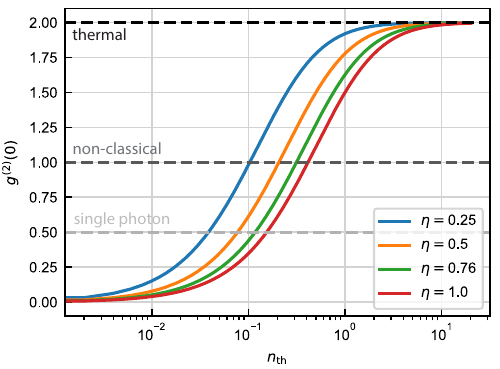} 
            	\caption{\textbf{Single-photon conversion in OMIT quantum interface.} Theoretical value of the autocorrelation function $g^{(2)}(0)$ of a single photon mapped in the phonon mode through the OMIT process with conversion efficiency $\eta$. The value $\eta=0.76$ corresponds to the conversion efficiency demonstrated in our OMC device. Horizontal dashed lines indicate the expected autocorrelation function for a thermal state $g^{(2)}(0)=2$, the criterion for non-classical phonon statistics $g^{(2)}(0)<1$, and the criterion for verification of a genuine single-phonon Fock state $g^{(2)}(0)<0.5$.}
            	\label{FigS1_single_photon_storage_criterion}
            \end{figure}

            Ultimately, OMIT-based photon-phonon interfaces may allow the conversion of genuine quantum states in long-lived mechanical modes. To assess the requirements in terms of conversion efficiency and thermal noise of the interface, we theoretically investigate the HBT measurement on a single photon state mapped in the mechanical mode $\hat{b}$ with efficiency $\eta$. To incorporate thermal noise with mean thermal phonon occupancy $n_\text{th}$ into our model, we consider the single-photon mode $\hat{s}$ and thermal mode $\hat{\delta}$ as independent modes that are measured in the same detection mode $\hat{b}$, such that
            \begin{align}
                \label{singlephoton_storage_mechanics}
                \hat{b} = \hat{s} + \hat{\delta}.
            \end{align}
            We assume the single photon mode $\hat{s}$ to be in a mixed state given by the limited conversion efficiency $\eta$ as
            \begin{align}
                \rho_s &= (1 - \eta)\, |0\rangle \langle 0| + \eta\, |1\rangle \langle 1|.
            \end{align}
            The single photon and thermal modes have the respective properties
            \begin{align}
                \langle \hat{s} \rangle &= 0, \\
                \langle \hat{s}^\dagger \hat{s} \rangle &= \eta, \\
                \langle \hat{s}^\dagger \hat{s}^\dagger \hat{s} \hat{s} \rangle &= 0
            \end{align}
            and
            \begin{align}
                \langle \hat{\delta} \rangle &= 0, \\
                \langle \hat{\delta}^\dagger(t)\, \hat{\delta}(t) \rangle &= n_{\text{th}}, \\
                \langle \hat{\delta}^\dagger(t)\, \hat{\delta}(t+\tau) \rangle &= n_{\text{th}}\, g_{\text{th}}^{(1)}(\tau).
            \end{align}
            We calculate the resulting second-order intensity autocorrelation function as defined in Eq.~\eqref{g2_definition} of the field given by Eq.~\eqref{singlephoton_storage_mechanics}. We simplify the expression in the denominator through
            \begin{align}
                \langle \hat{b}^\dagger(t) \hat{b}(t) \rangle 
                &= \langle (\hat{s}^\dagger(t) + \hat{\delta}^\dagger(t))(\hat{s}(t) + \hat{\delta}(t)) \rangle \\[6pt]
                &= \langle \hat{s}^\dagger \hat{s} \rangle 
                + \langle \hat{\delta}^\dagger \hat{\delta} \rangle 
                + \langle \hat{s}^\dagger \hat{\delta} \rangle 
                + \langle \hat{\delta}^\dagger \hat{s} \rangle \\[6pt]
                &= \eta + n_{\text{th}}.\label{denominator_single}
            \end{align}
            For the numerator, we find
            \begin{align}
                \langle \hat{b}^\dagger(t) \hat{b}^\dagger(t+\tau) \hat{b}(t+\tau) \hat{b}(t) \rangle 
                &= \langle (\hat{s}^\dagger + \hat{\delta}^\dagger(t))(\hat{s}^\dagger + \hat{\delta}^\dagger(t+\tau))
                (\hat{s} + \hat{\delta}(t+\tau))(\hat{s} + \hat{\delta}(t)) \rangle \\[6pt]
                &= n_{\text{th}}^{2}\!\left[ 1 + \left| g_{\text{th}}^{(1)}(\tau) \right|^{2} \right] 
                + 2\eta\,n_{\text{th}} \left(1 + \Re\left[g_{\text{th}}^{(1)}(\tau)\right]\right)\label{numerator_single}
            \end{align}
            Combining the results of Eqs~\eqref{denominator_single} and~\eqref{numerator_single} with the definition in~\eqref{g2_definition}, we obtain
            \begin{align}
                g^{(2)}(\tau)
                &= \frac{
                    n_{\text{th}}^{2}\bigl[1 + \lvert g_{\text{th}}^{(1)}(\tau) \rvert^{2}\bigr]
                    + 2\eta\,n_{\text{th}} \left(1 + \Re\left[g_{\text{th}}^{(1)}(\tau)\right]\right)
                }{
                    (\eta + n_{\text{th}})^{2}
                }.
            \end{align}
            At $\tau=0$, $|g_{\text{th}}^{(1)}(0)| = 1$ so we find
            \begin{align}
                g^{(2)}(0) 
                &= \frac{2\,n_{\text{th}}^{2} + 4\,\eta\,n_{\text{th}}}{
                (\eta + n_{\text{th}})^{2}}. \label{g2_single_photon_thermal}
            \end{align}
            Figure~\ref{FigS1_single_photon_storage_criterion} shows the expected autocorrelation functions for a single photon state mapped in the phonon mode through an OMIT process with added thermal noise from Eq~\eqref{g2_single_photon_thermal} as a function of thermal noise $n_\text{th}$ for various values of the conversion efficiency $\eta$. Equation~\eqref{g2_single_photon_thermal} shows that for a conversion efficiency of $\eta=0.76$ as demonstrated by our OMC device, thermal noise of $n_\text{th}<0.32$ or $n_\text{th}<0.12$ is required to demonstrate mapping of a single-photon and verification of its non-classical phonon statistics with $g^{(2)}(0)<1$ or verification of a genuine single-phonon state with $g^{(2)}(0)<0.5$, respectively. Improvements of the conversion efficiency to the ideal value of $\eta=1$ slightly alleviate these bounds to $n_\text{th}<0.41$ and $n_\text{th}<0.16$, respectively.
    \subsection{Total OMIT conversion-and-retrieval efficiency}
        In main text Figure 5, we measured the overall conversion-retrieval efficiency of our optimal memory, after reading the phonons out via a long red-detuned retrieval pulse with anti-Stokes scattering probabilities $p_\mathrm{aS} \approx 1$. The retrieved phonons $n_\mathrm{retrieve}$ contain three parts: (1) coherent phonons retrieved from interface $n_\mathrm{coh}$; (2) thermal phonons from the control pulse $n_\mathrm{pre}$ which should follow Fig.5(a) ; (3) instantaneous thermal phonons from the readout pulse which add an offset $n_\mathrm{0}$ to the total number of retrieved phonons depending on the retrieval pulse length we use. To elaborate, 
         \begin{align}
                       n_\mathrm{ret}
                        &= n_\mathrm{coh} + n_\mathrm{pre} + n_\mathrm{0} \\[6pt]
                        &= n_\mathrm{wav}\cdot \eta_\mathrm{store} \cdot \eta_\mathrm{retrieve} \cdot e^{-\frac{t}{T_\mathrm{1}}} + n_\mathrm{th}\cdot \eta_\mathrm{retrieve} + n_\mathrm{0} \\[6pt]
                        &= n_\mathrm{wav}\cdot \eta_\mathrm{c}\cdot \eta_\mathrm{int} \cdot \eta_\mathrm{c} \cdot A_\mathrm{retrieve}\cdot e^{-\frac{t}{T_\mathrm{1}}} + n_\mathrm{th}\cdot \eta_\mathrm{c} \cdot A_\mathrm{retrieve} + n_\mathrm{0} 
                \label{retrieved_phonon_expression}
                    \end{align}
        
         where we used $\eta_\mathrm{store} = \eta_\mathrm{c}\cdot \eta_\mathrm{int}$ if the temporal-mode overlap factor is $1$ in Equation~\ref{temporal-mode overlapping factor}; $\eta_\mathrm{c} = \kappa_\mathrm{e}/\kappa$ is the external coupling efficiency of the optical cavity, and $\eta_\mathrm{int} = 4\Gamma_{\text{eff}}\Gamma_{\text{sig}}/(\Gamma_{\text{eff}} + \Gamma_{\text{sig}})^2 \approx 1$ when the OMIT conversion process is perfectly bandwidth-matched. Moreover, we define the retrieval efficiency to be $\eta_\mathrm{retrieve} = \eta_\mathrm{c} \cdot A_\mathrm{retrieve}$ where we assume the anti-Stokes scattering probability of the readout pulse is $1$. $A_\mathrm{retrieve}$ represents the deviation from the perfect retrieval efficiency due to imperfect experimental conditions such as phonon decay occurring during the pulse and reduction of the readout anti-Stokes scattering probability because of shifts of the optical cavity resonance frequency commonly observed in OMC devices in response to strong optical pulses~\cite{chan_laser_2011,chen_bandwidth_tunable_2024}. Thermal phonons from prepulse follows the trend in Fig.5(a) and already contain the $ e^{-\frac{t}{T_\mathrm{1}}}$ decay. Based on the measurement in Fig.5(a) as well as signal input in the waveguide $n_\mathrm{wav}$, we can fit to obtain the overall conversion-retrieval efficiency. By fitting the data points to the expected phonon number, we obtain $A_\mathrm{retrieve} = 0.6$, thus the overall conversion-retrieval efficiency reads:
         \begin{align}
          \eta_\mathrm{cr} 
                   &= \eta_\mathrm{store} \cdot \eta_\mathrm{retrieve}\\[6pt]
                   &= \eta_\mathrm{c} \cdot \eta_\mathrm{int} \cdot \eta_\mathrm{c} \cdot A_\mathrm{retrieve}\\[6pt]
                   &=0.76\cdot 1 \cdot 0.76 \cdot 0.6\\[6pt]
                   &= 0.35
         \end{align}
        slightly lower than the expected overall efficiency of $\eta_\mathrm{c}^2 = 0.58$ due to the imperfect experimental conditions discussed above.
\section{Extended data of bandwidth matching measurements}
\begin{figure}
        \includegraphics[width = 1.0 \linewidth]{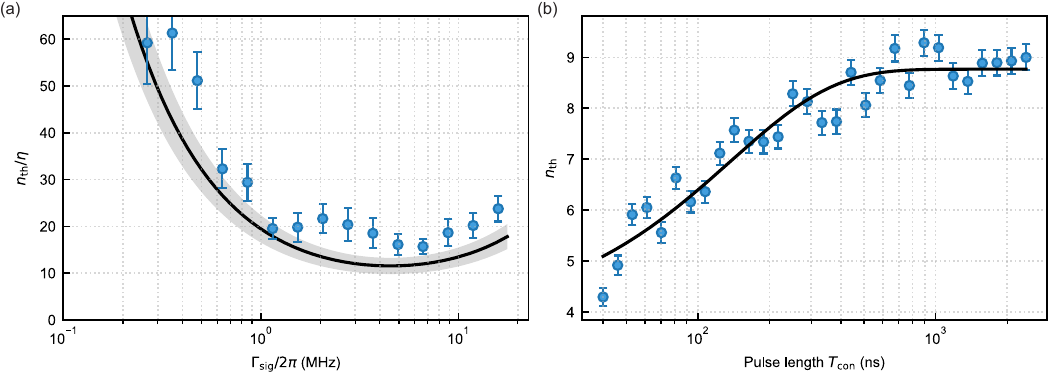} 
        \caption{\textbf{Bandwidth matching with higher control pulse power.} (a) Input-referred thermal noise measured as a function of signal bandwidth $\Gamma_\mathrm{sig}$. The solid line corresponds to the theoretical model discussed in the main text, where the added thermal noise $n_\mathrm{th}=8.77$ from the strong control pulse with pulse length $T_\mathrm{con}=\qty{2.4}{\micro s}$ and intracavity photon number $n_\mathrm{c}=7,800$ ($\Gamma_\mathrm{eff}/2\pi=\qty{4.5}{MHz}$) is independently calibrated (see (b)) and the shaded area shows the uncertainty of the theoretical model due to a relative error of $0.15$ in the calibration of $n_\mathrm{th}$. The conversion efficiency is assumed to be the ideal conversion efficiency expected from the cavity-impedance ratio $\eta=\eta_\mathrm{c}$. (b) Added thermal phonon noise $n_\mathrm{th}$ on the mapped mechanical state for the pulse configurations used in the measurement in (a) as a function of control pulse length $T_\mathrm{con}$. The solid black line is a fit to a phenomenological model of the heating process $n_\mathrm{th}=n_\mathrm{offset} + A [1-\exp(-T_\mathrm{con}/\tau_\mathrm{heat})]$.}
        \label{FigS4_bandwidth_matching_higher_power}
    \end{figure}

    In the main text, we performed the optimization of the signal pulse bandwidth $\Gamma_\mathrm{sig}$ at fixed intracavity photon number of the control pulse $n_\mathrm{c}=3,200$  and correspondingly fixed OMIT bandwidth of $\Gamma_\mathrm{eff}/2\pi=\qty{1.9}{MHz}$. To demonstrate the capability of our optomechanical interface to operate at larger bandwidths, we perform the same measurement at increased intracavity photon number $n_\mathrm{c}=7,800$ ($\Gamma_\mathrm{eff}/2\pi=\qty{4.5}{MHz}$) as shown in Fig.~\ref{FigS4_bandwidth_matching_higher_power}. The solid line corresponds to the theoretical model of the bandwidth matching condition in the OMIT conversion process introduced in the main text. As expected, compared to the data for smaller $\Gamma_\mathrm{eff}$ discussed in the main text, we observe a shift of the optimal $\Gamma_\mathrm{sig}$ to minimize the input-referred added noise $n_\mathrm{th}/\eta$ towards larger values of $\Gamma_\mathrm{sig}$, in good agreement with the model.
    
    We independently calibrate the added thermal noise induced by the strong OMIT control pulse using optomechanical sideband thermometry (see Fig.~\ref{FigS4_bandwidth_matching_higher_power}(b)). The saturation value of the added thermal phonon noise is set by a balance between the optical absorption heating effect and the optomechanical sideband cooling of the mechanical mode induced by the red-detuned control pulse with cooling rate $\Gamma_\mathrm{eff}\approx\Gamma_\mathrm{opt}=4n_\mathrm{c} g_0^2/\kappa$. Interestingly, we find no significant difference between the maximum added thermal noise $n_\mathrm{th}$ induced when using long control pulse lengths for different intracavity photon numbers of the control pulse ($n_\mathrm{th}=8.96$ for $n_\mathrm{c}=3,200$ and $n_\mathrm{th}=8.77$ for $n_\mathrm{c}=7,800$). This observation implies that, similar to the cooling rate, the optical absorption heating rate has an approximately linear scaling with intracavity photon number with $\Gamma_\mathrm{heat}\propto n_\mathrm{c}$, so that the saturation value of $n_\mathrm{th}$ is independent of the intracavity photon number, in agreement with previous studies that find a near-linear optical power dependence of the optical absorption heating in silicon OMCs~\cite{meenehan_pulsed_2015}.

\end{document}